\documentclass[preprint,aps ,nofootinbib]{revtex4}
\usepackage[colorlinks=true,linkcolor=blue,urlcolor=blue,filecolor=black,citecolor=red,pdfstartview=FitV,pdftitle={},pdfsubject={},pdfkeywords={},pdfpagemode=None,bookmarksopen=true]{hyperref}
\usepackage{graphicx}%Include figure filespp
\usepackage{epstopdf}%
\usepackage{amsmath}
\usepackage{amsfonts}
\usepackage{amssymb}
\usepackage{color}%
\usepackage{dcolumn}% Align table columns on decimal pointppp
\usepackage{slashed}
\usepackage{amssymb,ulem}
\usepackage{float}
\providecommand{\U}[1]{\protect\rule{.1in}{.1in}}

\newcommand{\f}{\begin{equation}}
\newcommand{\ff}{\end{equation}}
\newcommand{\fa}{\begin{eqnarray}}
\newcommand{\ffa}{\end{eqnarray}}

\newcommand{\blue}{\textcolor[rgb]{0.33,0.33,1.00}}

\begin{document}

\title{Informational properties of holographic Lifshitz field theory}
\author{Huajie Gong $^{1}$}
\thanks{huajiegong@qq.com}
\author{Peng Liu $^{2}$}
\thanks{phylp@jnu.edu.cn}
\author{Guoyang Fu$^{1}$}
\thanks{FuguoyangEDU@163.com}
\author{Xiao-Mei Kuang$^{1}$}
\thanks{xmeikuang@yzu.edu.cn}
\author{Jian-Pin Wu $^{1}$}
\thanks{jianpinwu@yzu.edu.cn}
\affiliation{
  $^{1}$ Center for Gravitation and Cosmology,
  College of Physical Science and Technology,
  Yangzhou University, Yangzhou 225009, China \ \\
  $^{2}$ Department of Physics and Siyuan Laboratory, Jinan University, Guangzhou 510632, China\\
}

\begin{abstract}

  In this paper, we explore the properties of holographic entanglement entropy (HEE), mutual information (MI) and entanglement of purification (EoP) in holographic Lifshitz theory. These informational quantities exhibit some universal properties of holographic dual field theory. For most configuration parameters and temperatures, these informational quantities change monotonously with the Lifshitz dynamical critical exponent $z$. However, we also observe some non-monotonic behaviors for these informational quantities in some specific spaces of configuration parameters and temperatures. A particularly interesting phenomenon is that a dome-shaped diagram emerges in the behavior of MI vs $z$, and correspondingly a trapezoid-shaped profile appears in that of EoP vs $z$. This means that for some specific configuration parameters and temperatures, the system measured in terms of MI and EoP is entangled only in a certain intermediate range of $z$.

\end{abstract} \maketitle

%%%
\section{Introduction}

The interplay between quantum information theory and quantum gravity has gained increasing attention during the last decade.
One central point of interest is entanglement measures in the context of the AdS/CFT correspondence \cite{Maldacena:1997re,Gubser:1998bc,Witten:1998qj,Aharony:1999ti}.

There are different ways to characterize different aspects of entanglement for quantum systems.
One particular measure is the entanglement entropy (EE).
Consider the physical system described by a density matrix $\rho$ characterizing the state of the system, which consists of two subsystems $A$ and $B$. Then the EE of subsystem $A$ with the reduced density matrix $\rho_A$
is just the von Neumann entropy for $\rho_A$. It is defined as $S_A=-Tr(\rho_A\log\rho_A)$ with $\rho_A=Tr_B\rho$.

In a holographic framework, termed the holographic entanglement entropy (HEE), EE has a simple geometric description known as the Rangamani-Takayanagi (RT) formula \cite{Ryu:2006bv,Takayanagi:2012kg,Lewkowycz:2013nqa},
%%%
\fa
S_A = \frac{\texttt{Area}(\gamma_A)}{4 G_N}\,,
\ffa
where $G_N$ is the bulk Newton constant, and $\gamma_A$ is the codimension$-2$ minimal surface in bulk geometry,
anchored to the asymptotic boundary such that $\partial\gamma_A=\partial A$.
The RT formula has also been extended to the covariant case, which is dubbed the Hubeny-Rangamani-Takayanagi (HRT) formula \cite{Hubeny:2007xt,Dong:2016hjy}. An important application of HEE is to study phase transitions; see for example Refs.\cite{Ling:2015dma,Ling:2016wyr,Ling:2016dck,Pakman:2008ui,Kuang:2014kha,Klebanov:2007ws,Zhang:2016rcm,Zeng:2016fsb,Albash:2012pd,Arefeva:2020uec,Liu:2020blk}.
Especially, in contrast to thermal entropy, EE is nonvanishing in the limit of zero temperature and thus it is an effective probe of quantum phase transitions (QPT) \cite{Ling:2015dma,Ling:2016wyr,Ling:2016dck,Pakman:2008ui,Klebanov:2007ws}.

We are also interested in the common information between two systems, which can be described by the mutual information (MI) \cite{Nielsen:QCQI}. Supposing that $A$ and $C$ are two disjoint entangling regions separated by a subsystem $B$, the MI is defined as
%%%%p
\fa
\label{MI-def}
I(A:C)=S_A+S_C-S_{A\cup C}\,.
\ffa
%%%%%%%%%p
In the above definition, the joint information of $A$ and $C$, $S_{A\cup C}$, is subtracted, which guarantees that we indeed obtain the common information of $A$ and $C$. Notice that a nontrivial MI requires that $S_{A\cup C}=S_B+S_{A\cup B\cup C}$.
In addition, the subadditivity guarantees the positive definiteness of MI.

MI has also been intensely studied in holography \cite{Casini:2004bw,Wolf:2007tdq,Headrick:2010zt}. Notice that EE is UV-divergent and needs to be regulated \cite{Bombelli:1986,Srednicki:1993im}. However, it is found that holographic MI (HMI) can remove the UV divergence of EE \cite{Casini:2004bw,Wolf:2007tdq,Headrick:2010zt}. Also, MI can partly cancel out the thermal entropy contribution \cite{Fischler:2012uv}. Therefore, MI is also an important quantity in holograpy.

When the system is in a pure quantum state, EE is a good quantum entanglement measure.
However, for mixed quantum states, EE is no longer a good measure of quantum entanglement because it is also sensible to classical correlations. Besides, MI is certain combination of EE such that it is not a genuinely new definition either in holography or in quantum information theory.

Entanglement of purification (EoP), $E_p$, is a candidate quantity measuring the correlation for \blue{a} bipartite mixed state $\rho_{AB}$ acting on $\mathcal{H}_{AB}\equiv \mathcal{H}_A\otimes\mathcal{H}_B$ \cite{Terhal:0202044,Bagchi:1502}. The EoP is defined as
\fa
E_p(\rho_{AB}):=\min_{|\psi\rangle_{AA'BB'}}S_{AA'}\,,
\ffa
%%%p
where we minimize over purification $\rho_{AB}=Tr[|\psi\rangle\langle\psi|_{AA'BB'}]$.
We would like to point out that EoP is a measure of correlations in terms of the entanglement of a pure state \cite{Terhal:0202044}. Also, notice that it represents the minimal value of quantum entanglement between $AA'$ and $BB'$ in an optimally purified system \cite{Terhal:0202044}. For simplicity, we also denote $E_p(\rho_{AB})=E_p(A:B)$ in what follows.

There are some simple properties for $E_p$ \cite{Terhal:0202044,Bagchi:1502}. First of all, if $\rho_{AB}$ is a pure state, i.e., $\rho_{AB}=|\psi\rangle\langle\psi|_{AB}$, $E_p$ reduces to the EE, i.e., $E_p(A:B)=S_A=S_B$. This means that no purification is needed. Secondly, $E_p$ vanishes if and only if $\rho_{AB}$ is uncorrelated, i.e., $\rho_{AB}=\rho_A\otimes\rho_B$.
More generally, we have the following inequalities for $E_p$ \cite{Terhal:0202044,Bagchi:1502}:
%%%%%p
\fa
&&
\label{EoPineq-v1}
\min(S_A,S_B)\geq E_p(A:B)\geq \frac{1}{2}I(A:B)\,,
\
\\
&&
\label{EoPineq-v2}
E_p(A:BC)\geq E_p(A:B)\,,
\
\\
&&
\label{EoPineq-v3}
E_p(AB:C)\geq \frac{1}{2}(I(A:C)+I(B:C))\,.
\ffa
%%%%%

In holography, it is proposed that the EoP is dual to the minimal area of the entanglement wedge cross section (EWCS) $E_w$ \cite{Takayanagi:2017knl,Nguyen:2017yqw}. Most features of $E_w$ match very well with those of $E_p$ in quantum field theory (QFT) \cite{Takayanagi:2017knl,Nguyen:2017yqw,Bao:2018gck,Umemoto:2018jpc}, which enhances the reliability of this prescription.

However, those early works mainly focused on the case of AdS$_3$. Recently, some pioneer works have already been devoted to studying the features of $E_p$ and its evolution behavior beyond AdS$_3$ by numerics \cite{Yang:2018gfq,Liu:2019qje,Huang:2019zph,Fu:2020oep}. In particular, an algorithm calculating $E_p$ for asymmetric configurations has been proposed for general holographic systems with homogeneity in Ref.\cite{Liu:2019qje} such that we can study the configuration-dependent characteristics of $E_p$. These studies indicate that in holography MI and EoP could have different abilities in depicting mixed state entanglement \cite{Huang:2019zph,Fu:2020oep}, which deserves further exploration. In addition, a similar concept of holographic complexity of purification (CoP) was also proposed in Ref.\cite{Ghodrati:2019hnn} in which the connection between holographic EoP and CoP was studied.

A notable feature for condensed matter systems is that many of them possess Lifshitz scaling symmetry as
%%%%p
\fa
\label{Lifshitz-scaling}
t\rightarrow\lambda^z t,\,~~~~~~\vec{x}\rightarrow\lambda\vec{x}\,,
\ffa
%%%%%%%%%%%
where $z$ is the Lifshitz exponent. When $z>1$, the isotropy between time and space is broken. The dual boundary field theory flows to a non-relativistic fixed point, which possesses Lifshitz symmetry. When $z=1$ the dual boundary field theory flows to a relativistic fixed point. In holography, the gravity descriptions of Lifshitz fixed points have been obtained in Ref. \cite{Hertz:1976zz}.  Many Lifshitz black hole geometries have also been implemented in Refs. \cite{Kachru:2008yh,Danielsson:2009gi,Mann:2009yx,Bertoldi:2009vn,Taylor:2008tg,Pang:2009ad,Pang:2009pd,Balasubramanian:2009rx,AyonBeato:2009nh,Cai:2009ac,Myung:2009up,Dehghani:2011tx,Keranen:2012mx,Tarrio:2011de,Kuang:2017rpx}. Notice thar, here we call $z$ Lifshitz exponent to avoid confusion with the dynamical critical exponent near the phase transition critical point \cite{Hohenberg:1977}. In holographic theory, it is found that the dynamical critical exponent near the phase transition critical point may be independent of the geometry Lifshitz exponent $z$ \cite{Natsuume:2018yrg,Li:2019oyz}.

Recently, the informational quantities have been explored for holographic dual field theory with Lifshitz symmetry; see Refs.\cite{Chakraborty:2014lfa,Karar:2020cvz,Dong:2012se,Alishahiha:2015goa,Mishra:2018tzj} and references therein. However, most of these studies focused only on the HEE or MI in the background with zero charge density, and there have been few investigation on the informational quantities at finite density , especially the EoP. In this work, we shall study the related information quantities in holographic Lifshitz dual field theory with finite charge density.

This paper is organized as follows. We review the charged Lifshitz black brane and deduce the expressions of HEE, MI and EoP with this background in Section \ref{section-b} and Section \ref{section-c}, respectively. Then in Section \ref{section-d}, the numerical results of these informational-related quantities are presented and the corresponding properties are explored. Our results are summarized in Section \ref{section-e}.

%%%%%p
\section{Charged Lifshitz black brane}\label{section-b}

To have a holographic Lifshitz dual boundary field theory with finite charged density,
we consider the following Einstein-Maxwell-dilaton (EMD) action in four-dimensions bulk spacetimes \cite{Tarrio:2011de}
%%%%
\fa
\label{action-HV}
S=-\frac{1}{16\pi G}\int
d^{4}x\sqrt{-g}\Big[R-\frac{1}{2}(\partial\psi)^2+V(\psi)-\frac{1}{4}\Big(e^{\lambda_1\psi}F^2+e^{\lambda_2\psi}\mathcal{F}^2\Big)\Big]\,.
\ffa
The above action includes a dilaton field $\psi$ as well as two $U(1)$ gauge fields, $A$ and $\mathcal{A}$, with field strengths $F_{\mu\nu}$ and $\mathcal{F}_{\mu\nu}$, respectively.
$A$ is the real Maxwell field which sources the charge while $\mathcal{A}$ plays the role of an auxiliary field which supports an asymptotic Lifshitz geometry. $\lambda_1$, $\lambda_2$ are free parameters of the theory. The potential $V(\psi)=V_{0}e^{\gamma\psi}$.

The action \eqref{action-HV} supports a charged Lifshitz black brane solution:
\begin{eqnarray}
  &&
  \label{Metric}
  ds^{2}=-r^{2z}f(r)dt^2+\frac{dr^2}{r^2 f(r)}+r^2(dx^2+dy^2)\,,
  \
  \\
  &&
  \label{fr}
  f(r)=1-\frac{M}{r^{z+2}}+\frac{Q^2}{r^{2(z+1)}}\,,
  \
  \\
  &&
  \label{At}
  A_t=\mu r_h^{-z}\Big(1-\Big(\frac{r_h}{r}\Big)^{z}\Big)\,,
  \
  \\
  &&
  \label{cAt}
  \mathcal{A}_t=-\slashed{\mu}r_h^{2+z}\Big(1-\Big(\frac{r}{r_h}\Big)^{2+z}\Big)\,,
\end{eqnarray}
where $r_h$ is the position of the horizon, and $M$ and $Q$ are the mass and charge of the black brane, respectively.
The parameters $V_0$, $\lambda_1$ and $\lambda_2$ in the action could be determined in terms of $z$ as
%%%

\fa
\label{parameters}
&&
V_{0}=(z+1)(z+2)\,,
\nonumber
\\
&&
\lambda_1=\sqrt{\frac{2(z-1)}{2}}\,,
\nonumber
\\
&&
\lambda_2=-\frac{2}{\sqrt{z-1}}\,.
\ffa
Notice that $\gamma=0$ here.

The horizon condition $f(r_h)=0$ gives the relation between $M$ and $Q$,
\fa
r_h^{2(z+1)}-M r_h^{z}+Q^2=0\,.
\ffa

In terms of $Q$ and $z$, $\mu$ and $\slashed{\mu}$ can be expressed as
\fa
&&
\label{mu}
\mu=\frac{2Q}{\sqrt{z}}\,,
\
\\
&&
\label{mu-s}
\slashed{\mu}=\sqrt{\frac{2(z-1)}{2+z}}\,,
\ffa
where $\mu$ is the chemical potential of the dual boundary field theory.
The Hawking temperature can then be easily worked out as
\fa
\label{HT-v1}
\hat{T}=\frac{(2+z)r_h^z}{4\pi}\Big[1-\frac{z}{2+z}Q^2 r_h^{2(-z-1)}\Big]\,.
\ffa
%%%
For convenience, we can set $r_h=1$ by scaling symmetry.
Now, for a given $z$, this black brane solution is determined by the scaling-invariant quantity $T\equiv\hat{T}/\mu$.

In the limit of zero temperature, it is easy to find that the IR geometry of this black brane is $AdS_2\times \mathbb{R}_2$, which is the same as that of RN-AdS geometry. However, we note that the curvature radius of $AdS_2$ is $L_2=1/\sqrt{z(z+2)}$, which depends on the Lifshitz exponent $z$. For the detailed derivation, we can refer to Refs.\cite{Wu:2013xta,Wu:2014rqa}.

%%%%%
\section{Holographic Information-related Quantities}\label{section-c}

For the static case, the RT formula is still applicable to the Lifshitz geometry and the holographic calculation for EE matches the result for the Lifshitz field theory \cite{Solodukhin:2009sk,Nesterov:2010yi}. However, we would also like to point out that since the entanglement wedges do not naturally reach the boundary of the spacetime \cite{Gentle:2015cfp}, the construction of the covariant Lifshitz formula, i.e., the equivalent of the HRT formula, is not direct. In Refs.\cite{Cheyne:2017bis,Janiszewski:2017tas}, causal propagation of the high frequency Lifshitz modes is used to construct entanglement wedges, an alternative formula to HRT, and it is shown that some field theory results for the EE can be reproduced.

In this paper, the Lifshitz geometry we study is static. Therefore, we shall follow the RT formula to calculate the HEE, then the MI and EoP.
For the convenience of the numerical calculation, we transform the coordinates as $\rho=1/r$ such that the horizon is at $\rho=1$
and the boundary is at $\rho=0$. We re-express the black brane geometry as
\fa
ds^2=-\rho^{-2z}U(\rho)dt^2+\frac{1}{\rho^{2}U(\rho)}d\rho^2+\frac{1}{\rho^2}(dx^2+dy^2)
\ffa
with
\fa
U(\rho)=1-M\rho^{2+z}+Q^2\rho^{2(1+z)}.
\ffa
%%%p

%%%%%p
\begin{figure}[ht!]
  \centering
  \includegraphics[width=0.65\textwidth]{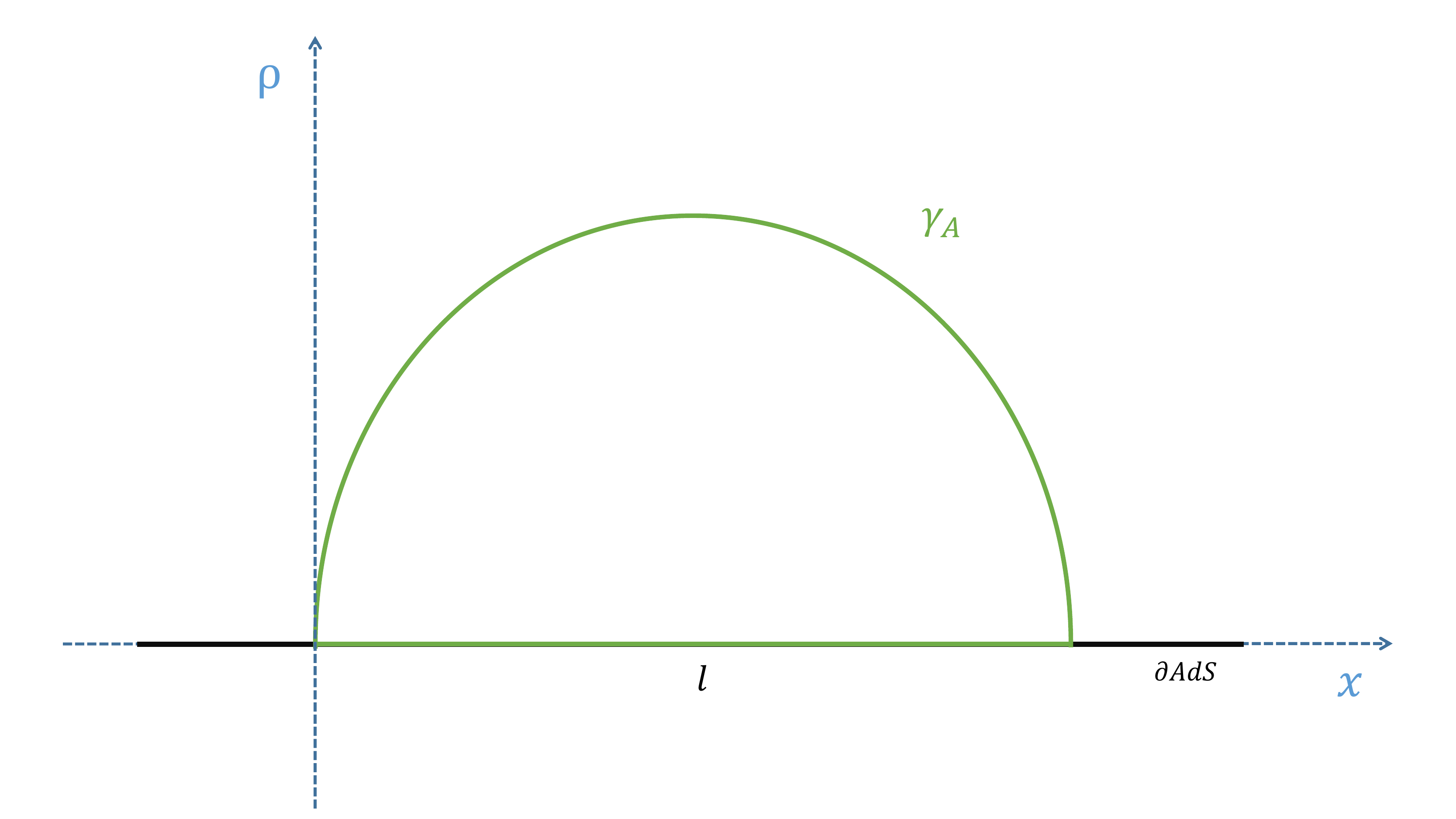}
  \caption{Cross-sectional view of an extreme surface $\gamma_A$ in bulk produced by the subsystem $A$ with infinite configuration in the boundary. This subsystem $A$ has width $l$ along the $x$ direction and is infinite along the $y$ direction.}
  \label{EE-cartoon}
\end{figure}
%%%%ppp
In this paper, we consider an infinite strip subsystem in the dual boundary, which can be specifically depicted as
$A:=\{0 < x < l, -\infty < y < \infty\}$ (see FIG.\ref{EE-cartoon}).
This setup preserves the translation invariance of the minimal surfaces along the $y$ direction
such that we can parametrize the minimal surface by the radial coordinate $\rho(x)$.
Then, we can write down the regularized HEE of the minimum surface and the corresponding width of the strip:
%%%%%%%%p
\fa
\label{HEE-S}
\hat{S}=2\int_{\epsilon}^{\rho_*}\frac{\rho_*^{2}}{\rho^{2}\sqrt{\rho_*^{4}-\rho^4}\sqrt{-M\rho^{z+2}+Q^2\rho^{2z+2}+1}}d\rho\,,
\ffa
\fa
\hat{l}=2\int_{\epsilon}^{\rho_*}\frac{\rho^{2}}{\sqrt{\rho_*^{4}-\rho^4}\sqrt{-M\rho^{z+2}+Q^2\rho^{2z+2}+1}}d\rho\,,
\ffa
%%%%%%%%
where $\rho_*$ is the location of the turning point of the minimum surface at which $\rho'(x)|_{\rho=\rho_{*}}=0$, and $\epsilon$ is the UV cutoff.
We are interested in the scaling-invariant HEE and width, which are $S\equiv\hat{S}/\mu$ and $l\equiv\hat{l}\mu$.

\begin{figure}[H]
  \centering
  \includegraphics[scale=0.5]{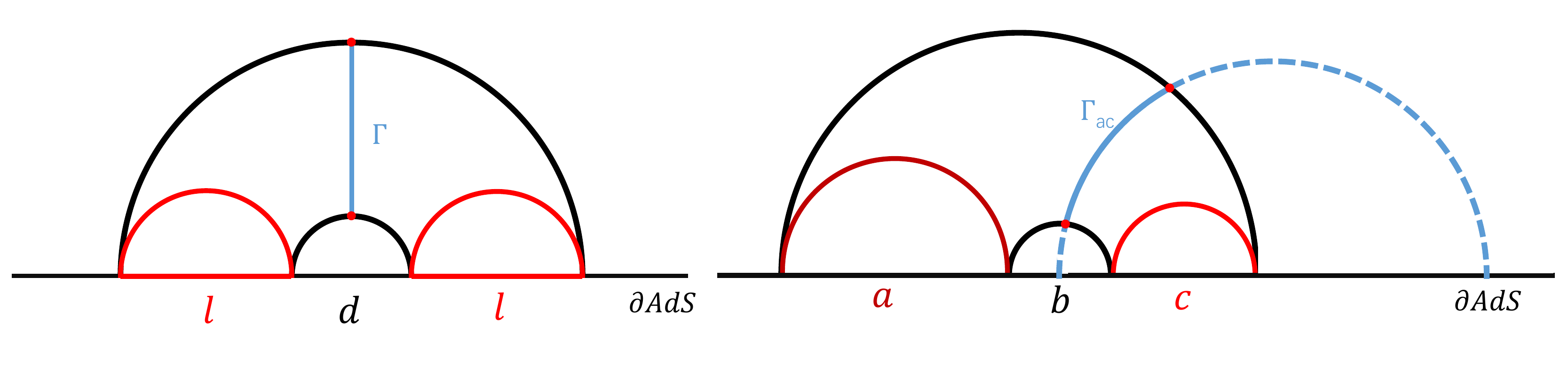}\
  \caption{Schematic configuration for computing MI and EoP for the symmetric case (left) and asymmetric case (right)). The two subsystems are separated by the region with width $b$ ($b\equiv d$ for the symmetric case). The red curves represent the area of the disconnected configuration and the black curves denote the area of connected configuration. MI is depicted by their difference. $E_p$ is calculated by the entanglement wedge $\Gamma$ shown by the blue line.
  }
  \label{EoPcartoonpicture}
\end{figure}

For MI, we also consider infinite stripe geometries along the $y$ direction.
We denote the widths of $A$, $B$ and $C$ along the $x$ direction as $a$, $b$ and $c$, respectively.
Once the HEE is worked out, MI can be calculated directly in terms of Eq.\eqref{MI-def}.
When $a=c$, the configuration is symmetric. For this case, we denote $a=c=l$ and $b=d$.
We show the schematic symmetric configuration for computing MI in FIG.\ref{EoPcartoonpicture}.
%%%%p

In holography, EoP $E_p$ is proposed as the area of the minimal EWCS $E_w$ for a connected configuration of MI, i.e., $E_p=E_w$, which is given by \cite{Takayanagi:2017knl,Nguyen:2017yqw}
%%%%p
\fa
E_p(\rho_{AC})=\min_{\Sigma_{AC}}\Big(\frac{Area(\sum_{AC})}{4G_{N}}\Big),\,
\ffa
where $\Sigma_{AC}$ is the cross-section in the entanglement wedge of $A\cup C$.
Then, we can explicitly derive the concrete expression of $E_p$ in our present model as
%%%%
\fa
E_p=\frac{1}{4G_N}\int_{{\Gamma}}
% {\rho_*^{(b)}}^{\rho_*^{(a+b+c)}}
\frac{\rho}{\sqrt{1-M\rho^{2+z}+Q^2\rho^{2(1+z)}}}d\rho
\ffa
%%%%
We show a schematic configuration for computing EoP in FIG.\ref{EoPcartoonpicture}.
Next, we follow the numerical procedure outlined in Ref.\cite{Liu:2019qje} to study the properties of HEE, MI and EoP in holographic Lifshitz dual field theory.

%%%
\section{Numerical results}\label{section-d}
%%%%%%
\subsection{Holographic Entanglement Entropy}\label{subsec:hee}

\begin{figure}[H]
  \center{
    \includegraphics[scale=0.5]{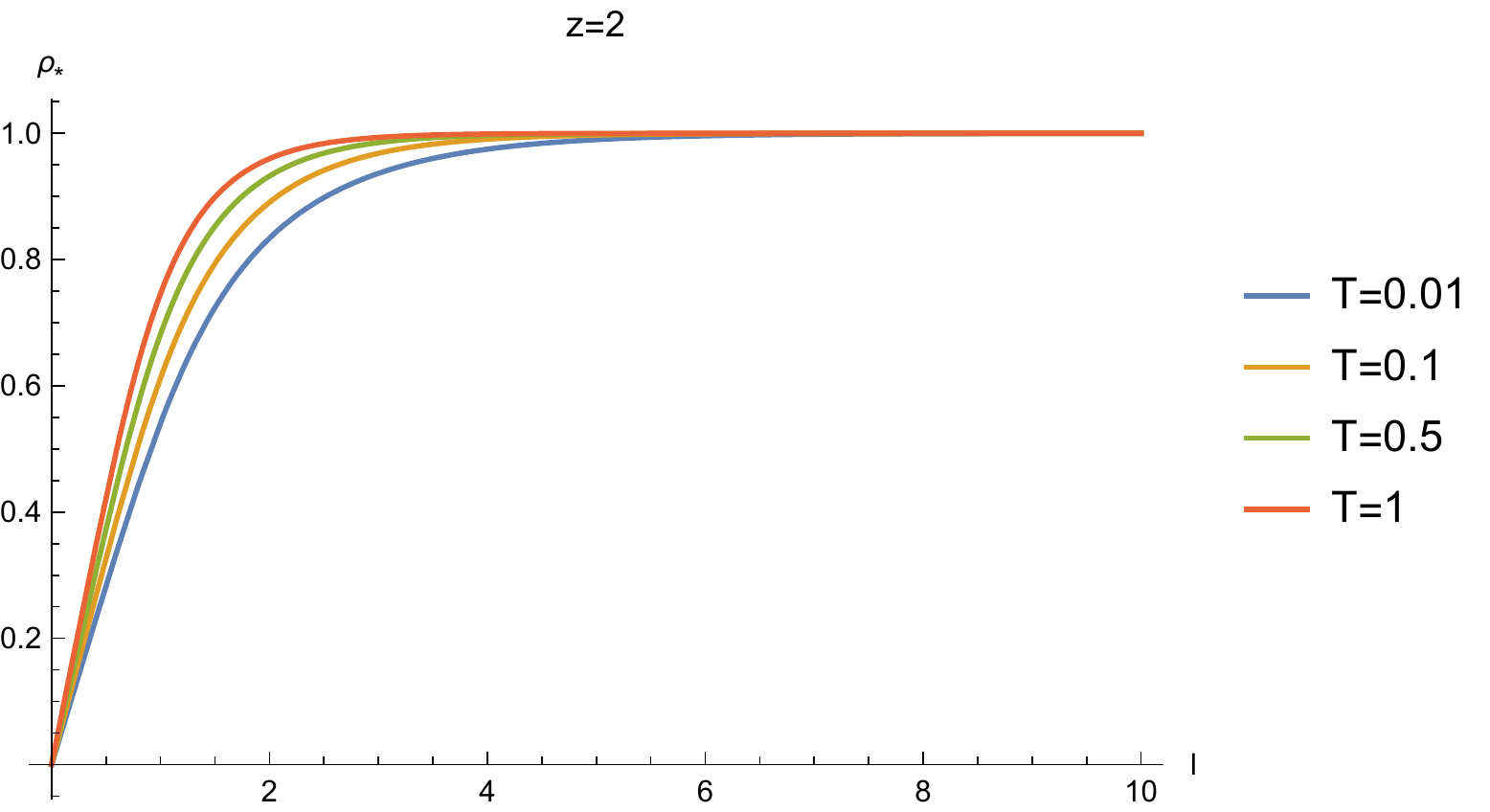}\ \hspace{0.5cm}
    \includegraphics[scale=0.5]{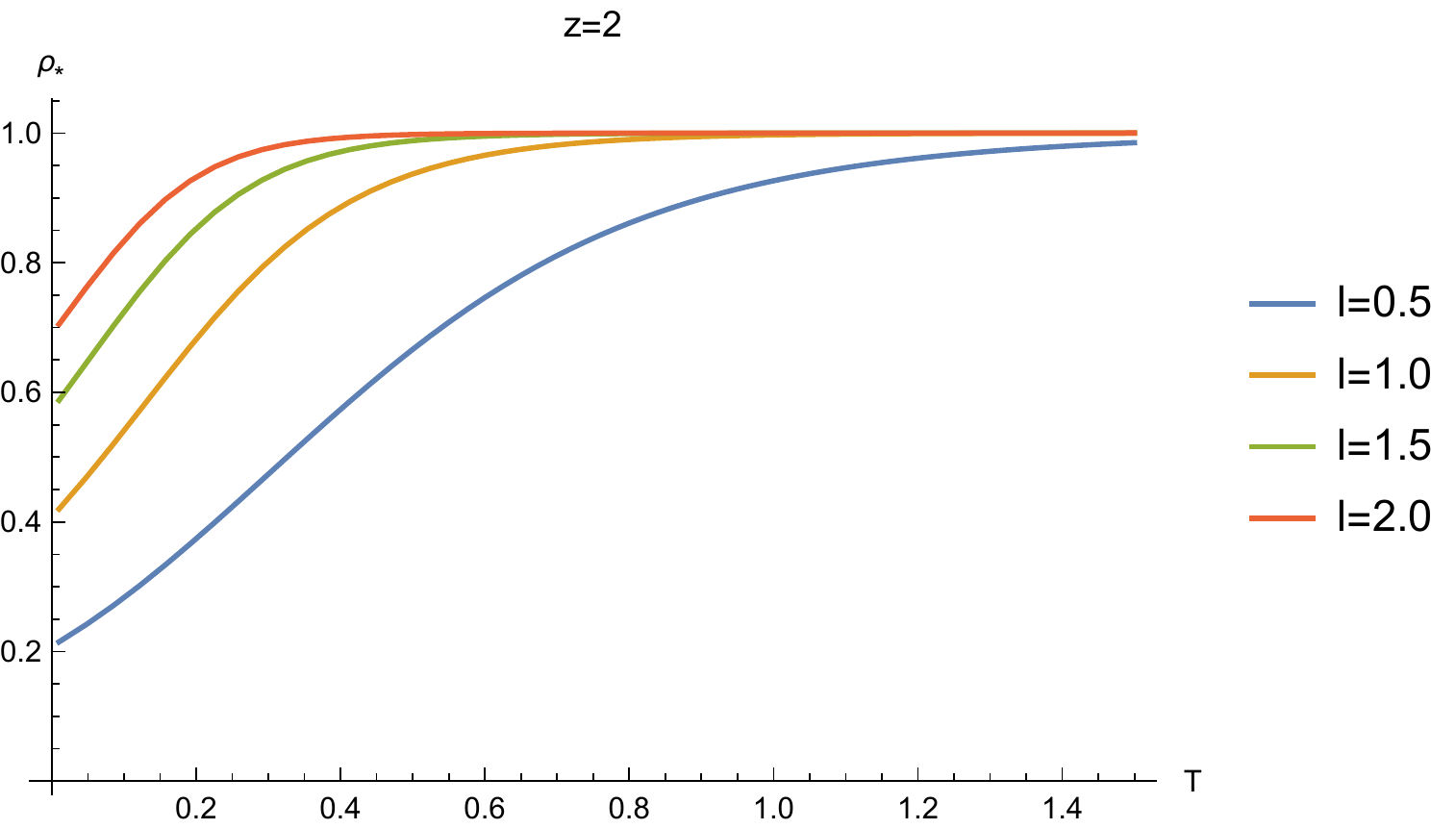}\
    \caption{\label{rhoz2} Variation of turning point $\rho_*$ with (left) width $l$ for different temperature $T$ and (right). $T$ for different $l$. Here $z=2$.
    }}
\end{figure}

We first explore the behaviors of the turning point $\rho_*$, which can provide some insights into the holographic informational quantities. Figure \ref{rhoz2} shows the turning point $\rho_*$ as a function of width $l$ for different temperatures (left) and as a function of temperature $T$ for different widths $l$ (right) for a charged Lifshitz black brane with $z=2$. It is easy to find that $\rho_*$ increases monotonically with $l$. In the limit of $l\rightarrow\infty$, it stretches to the horizon of the black brane, while in the limit of $l\rightarrow 0$, it shrinks to the AdS boundary, i.e., $\rho_*\rightarrow 0$. The behaviors are qualitatively similar to that of the RN-AdS background \cite{Liu:2019qje} and the Gubser-Rocha model \cite{Fu:2020oep}. However, notice that for the Gubser-Rocha model, there is a region of $l$ where $\rho_*$ is almost vanishing \cite{Fu:2020oep}. This is a peculiar property of the Gubser-Rocha model, different from the RN-AdS background and the charged Lifshitz geometry studied here.

\begin{figure}[H]
  \center{
    \includegraphics[scale=0.5]{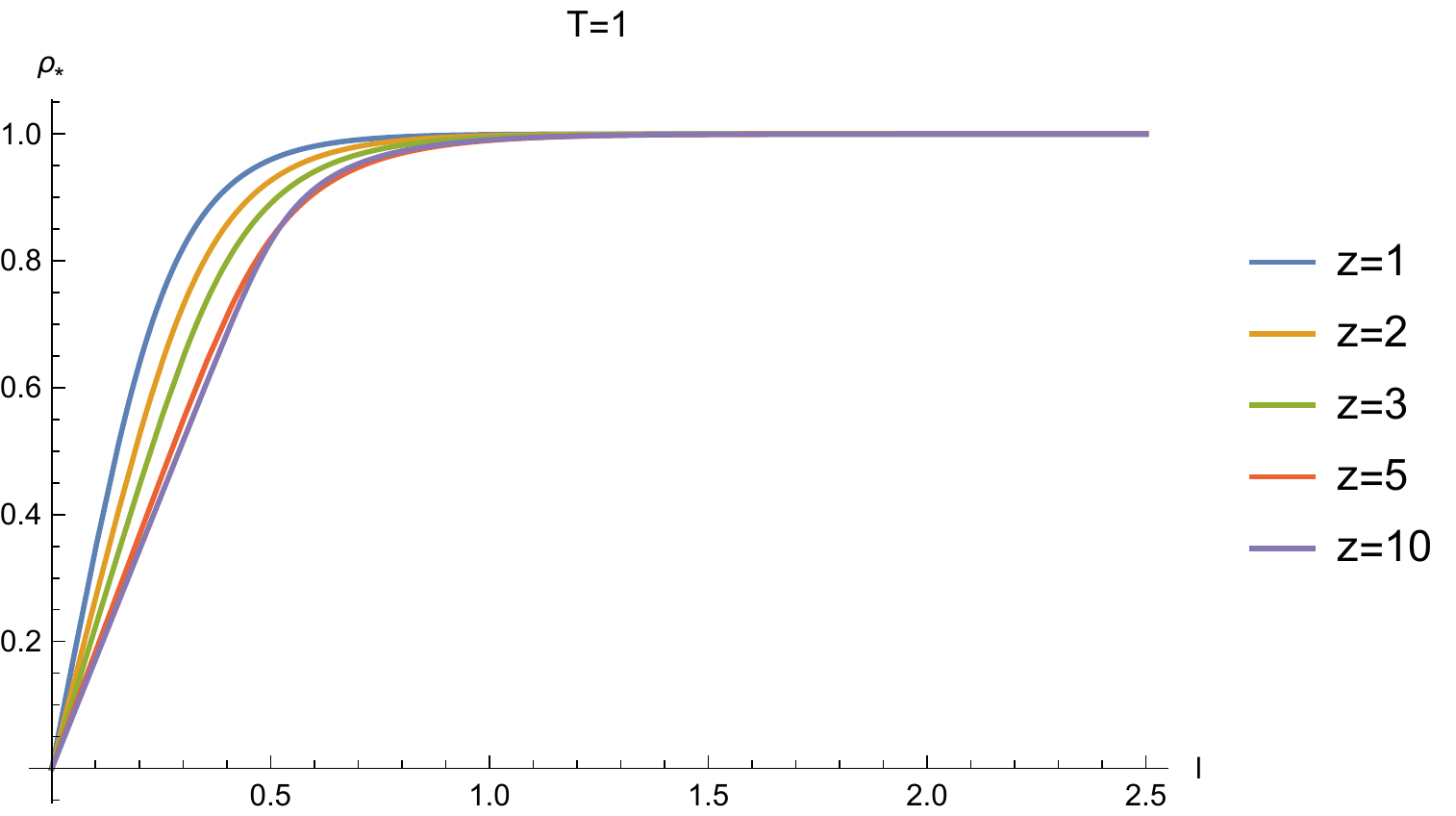}\ \hspace{0.5cm}
    \includegraphics[scale=0.5]{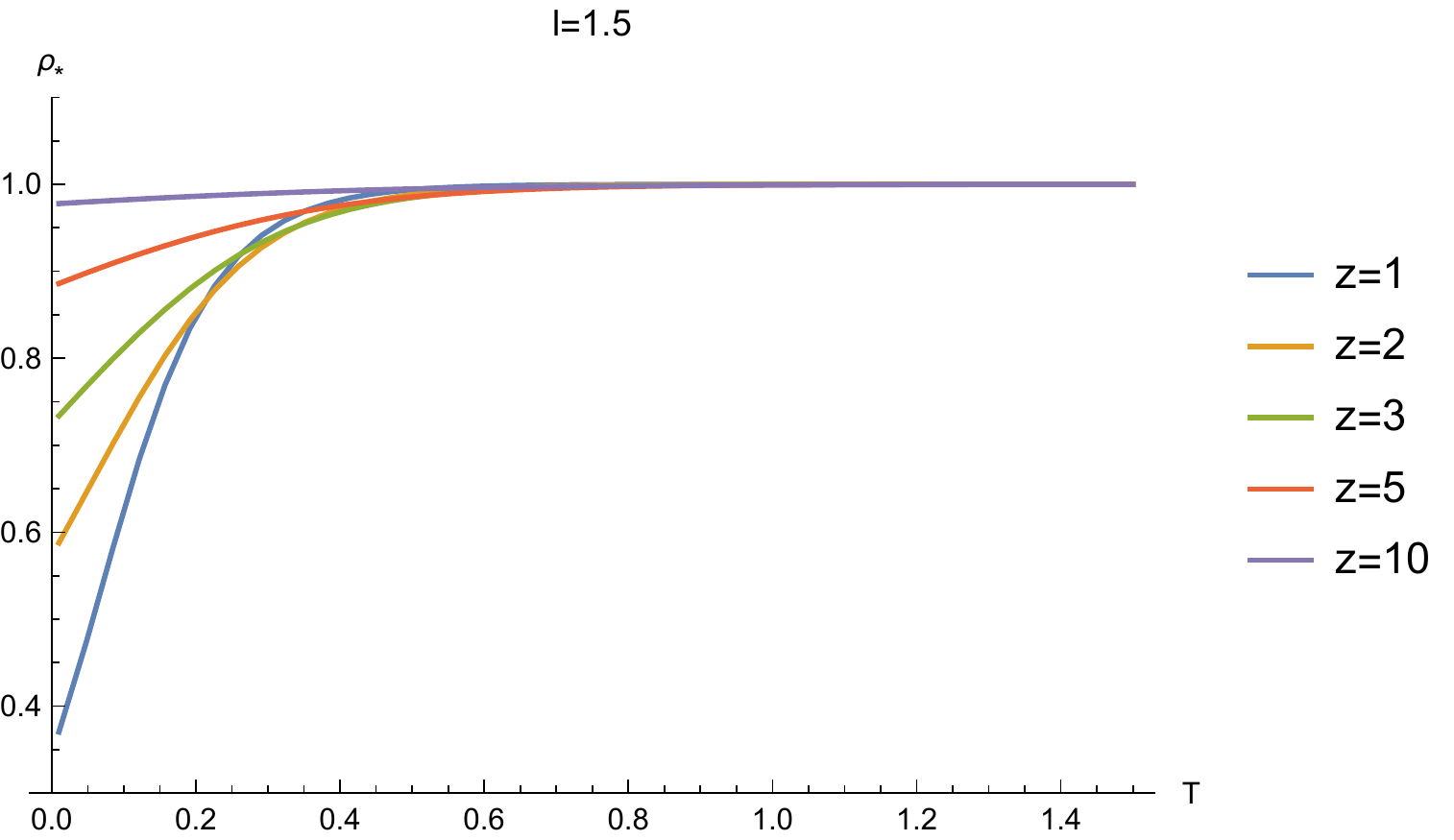}\\
    \caption{\label{rhodiffz} Width and temperature behaviors of the turning point $\rho_*$ for different $z$.
    }}
\end{figure}

For the temperature behavior of $\rho_*$, we see that $\rho_*$ increases with the temperature and approaches the black brane horizon in the high temperature limit (right-hand plot in FIG.\ref{rhoz2}).
The reason is that the horizon radius increases with the temperature and hence the minimum surface tends to approach the black hole horizon in the high temperature limit. However, we note that for finite $l$, $\rho_*$ is finite even in the limit of zero temperature. The underlying reason is that the geometry of the Lifshitz system in the zero temperature limit is regular. This behavior is similar to that of the RN-AdS background but is different from that of the Gubser-Rocha model.
Furthermore, in FIG.\ref{rhodiffz}, we show the width and the temperature behaviors of the turning point $\rho_*$ for different $z$. This confirms the observations of the behavior of $\rho_*$ for a charged Lifshitz black brane for $z=2$.
Therefore, we conclude that the qualitative behaviors of the turning point $\rho_*$ of the charged Lifshitz geometry are closely similar to those of the RN-AdS background. However, notice that the curves of $\rho_*$ for different $z$ in FIG.\ref{rhodiffz} intersect each other. This indicates non-monotonic behavior of $\rho_*$ with $z$ in some specific regions of $l$ and $T$. In addition, from the right-hand plot in FIG.\ref{rhodiffz}, we see that in the limit of zero temperature, the turning point stretches deeper into the bulk and is closer to the horizon of the black brane for larger $z$. This shows that HEE is determined by the near-horizon geometry at the zero temperature limit when $z$ is large.
%%%%%%%%%%%%%%%%p
%%%%%%%%%%%p
%%%%%%%%%%%%%%%%p
%%%%%%%%%%%p
%%%%%%%%%%%%%%%%p
\begin{figure}[H]
  \center{
    \includegraphics[scale=0.5]{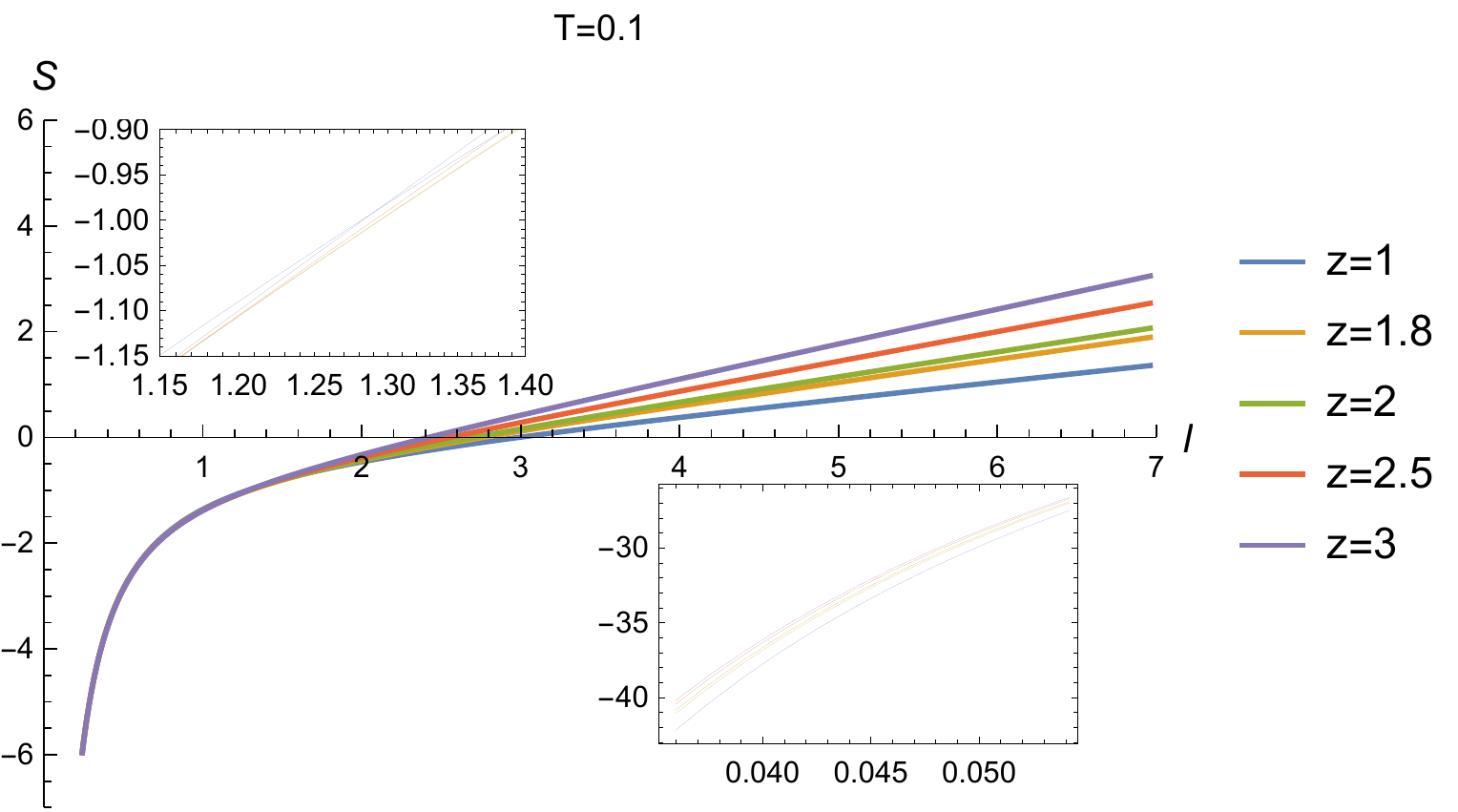}\ \hspace{0.5cm}
    \includegraphics[scale=0.5]{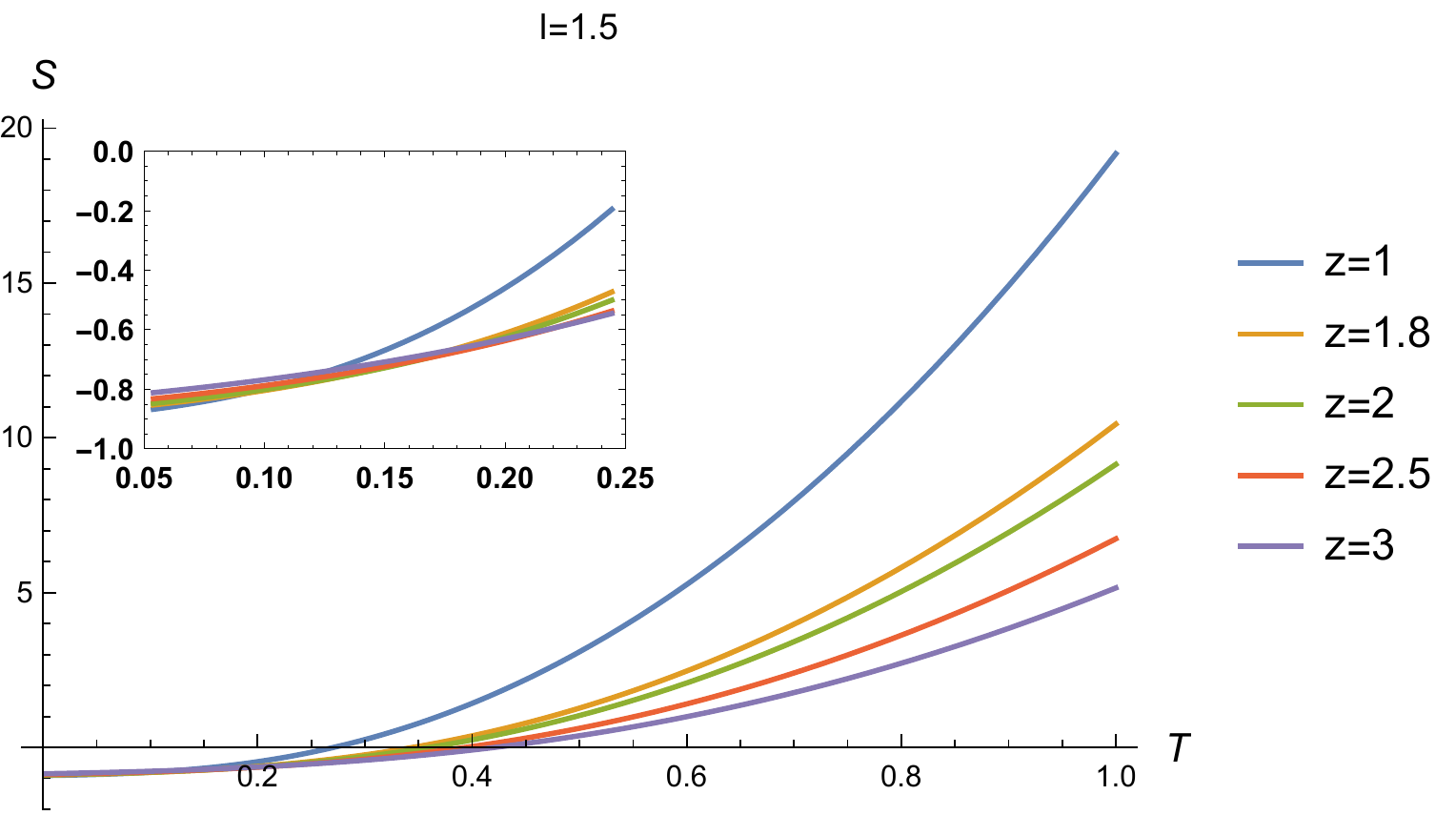}\
    \caption{\label{figHEE_differentz} (left) HEE vs $l$ for different $z$ ($T=0.1$) and (right) HEE vs $T$ for different $z$ ($l=1.5$).
    }}
\end{figure}
%%%%%%%%%%%p

  Next, we explore the behaviors of HEE for different $z$, as shown in FIG.\ref{figHEE_differentz}. We see that for different $z$, HEE decreases monotonically with decreasing $l$ and approaches negative infinity as $l\rightarrow 0$ (left). The reason is that as the subregion in consideration shrinks, the number of entangled degrees of freedom will also decrease.
As the temperature drops, HEE also decreases monotonically. This is partly due to the contribution from the thermal entropy, which has a monotonically increasing dependence on the temperature.
In the limit of zero temperature, HEE is finite (right). This behavior in the low temperature region is similar to that of the RN-AdS system \cite{Liu:2019qje}, but different from that of the Gubser-Rocha model studied in Ref. \cite{Fu:2020oep}, for which the HEE in the low temperature region exhibits a non-monotonic temperature behavior.

\begin{figure}[H]
  \center{
    \includegraphics[scale=0.5]{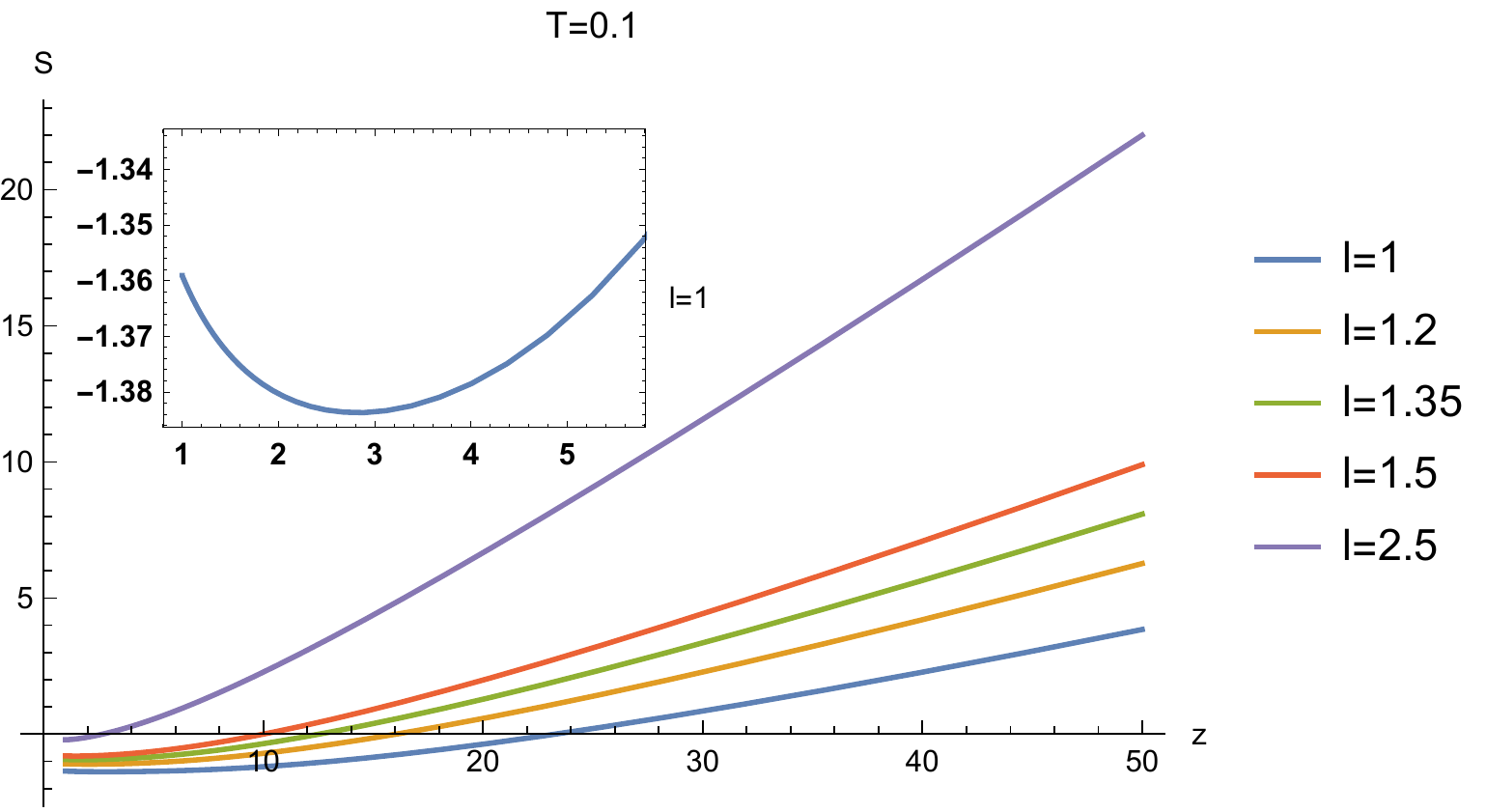}\ \hspace{0.5cm}
    \includegraphics[scale=0.5]{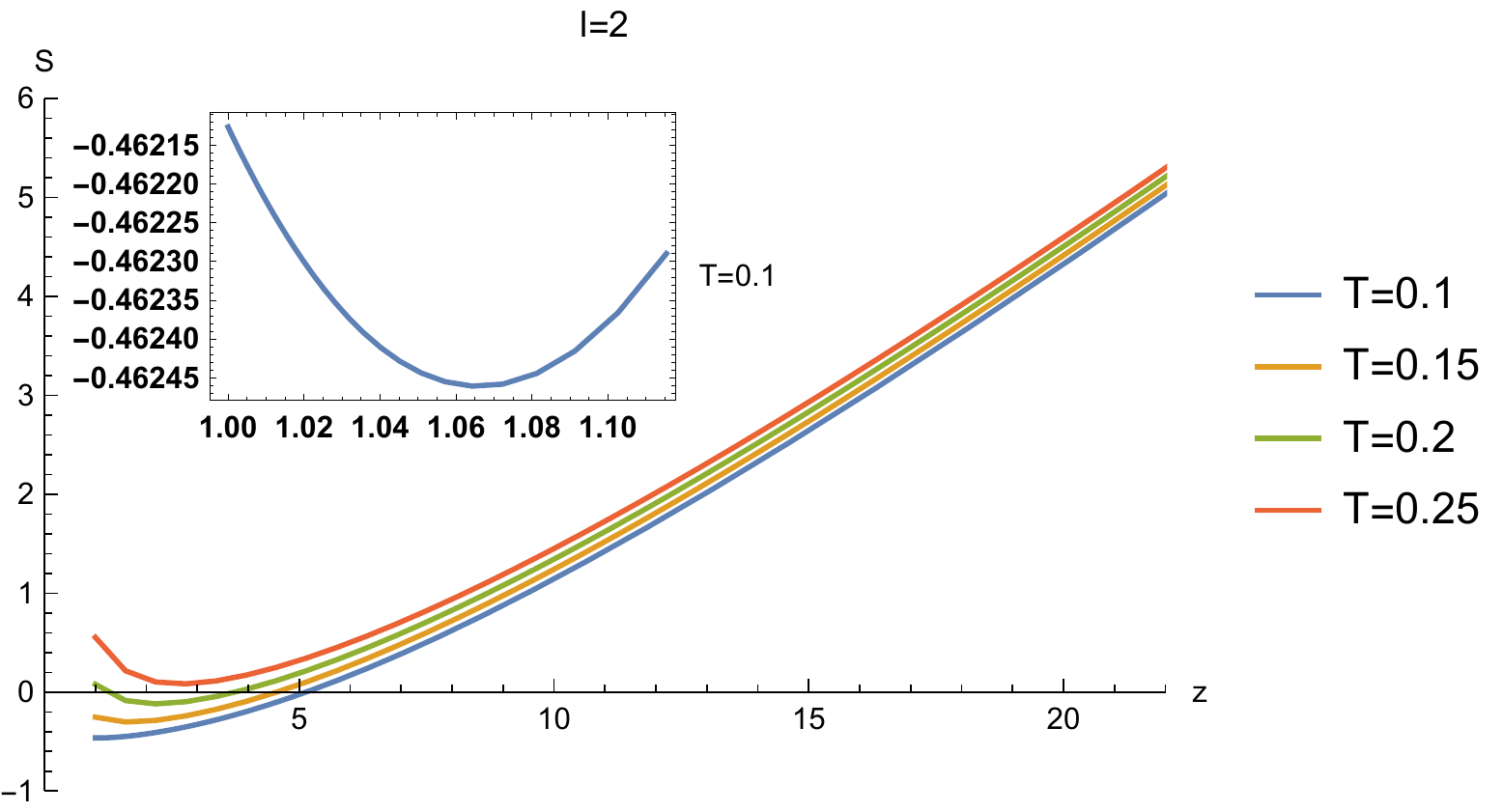}\ \\
    \includegraphics[scale=0.5]{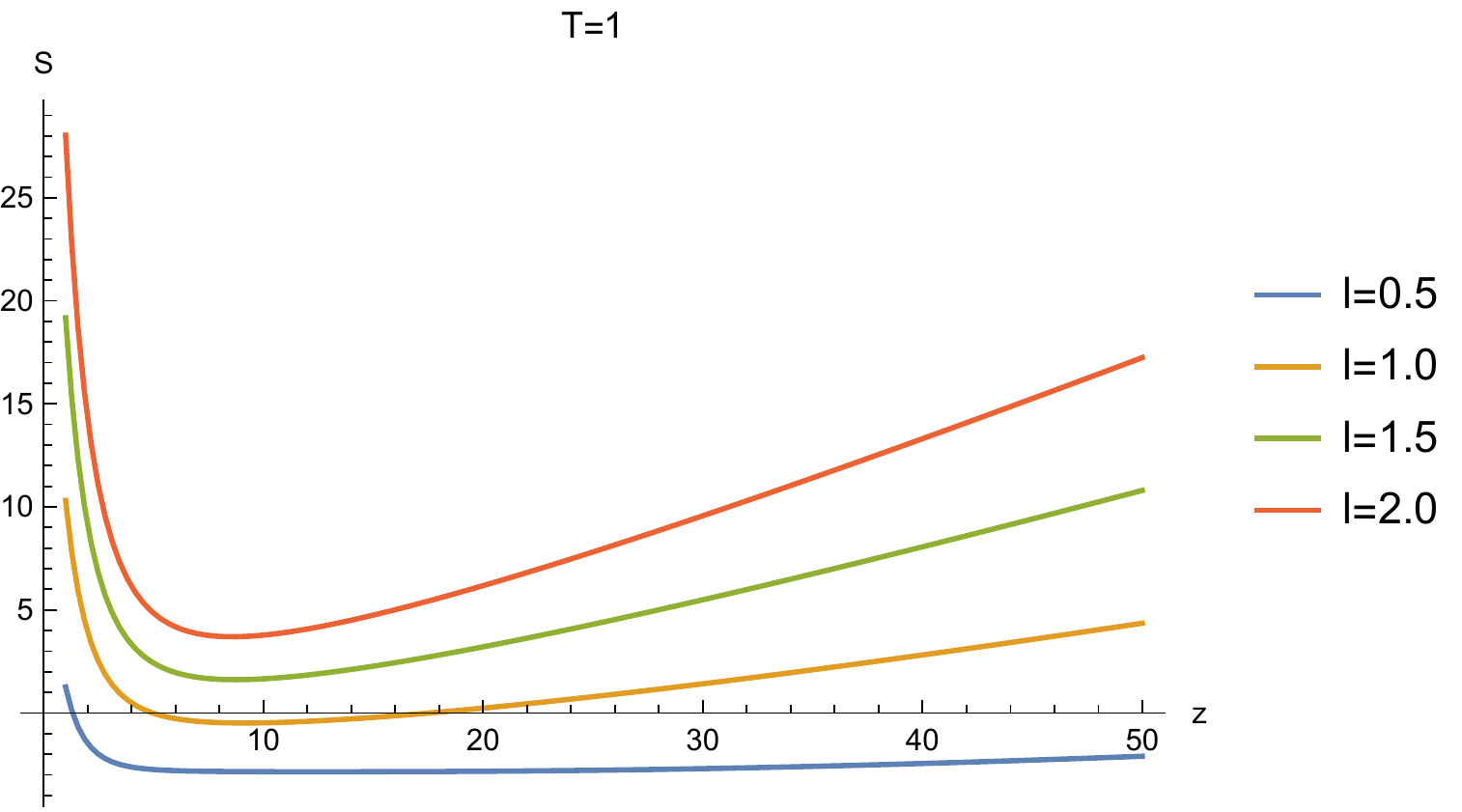}\ \hspace{0.5cm}
    \includegraphics[scale=0.5]{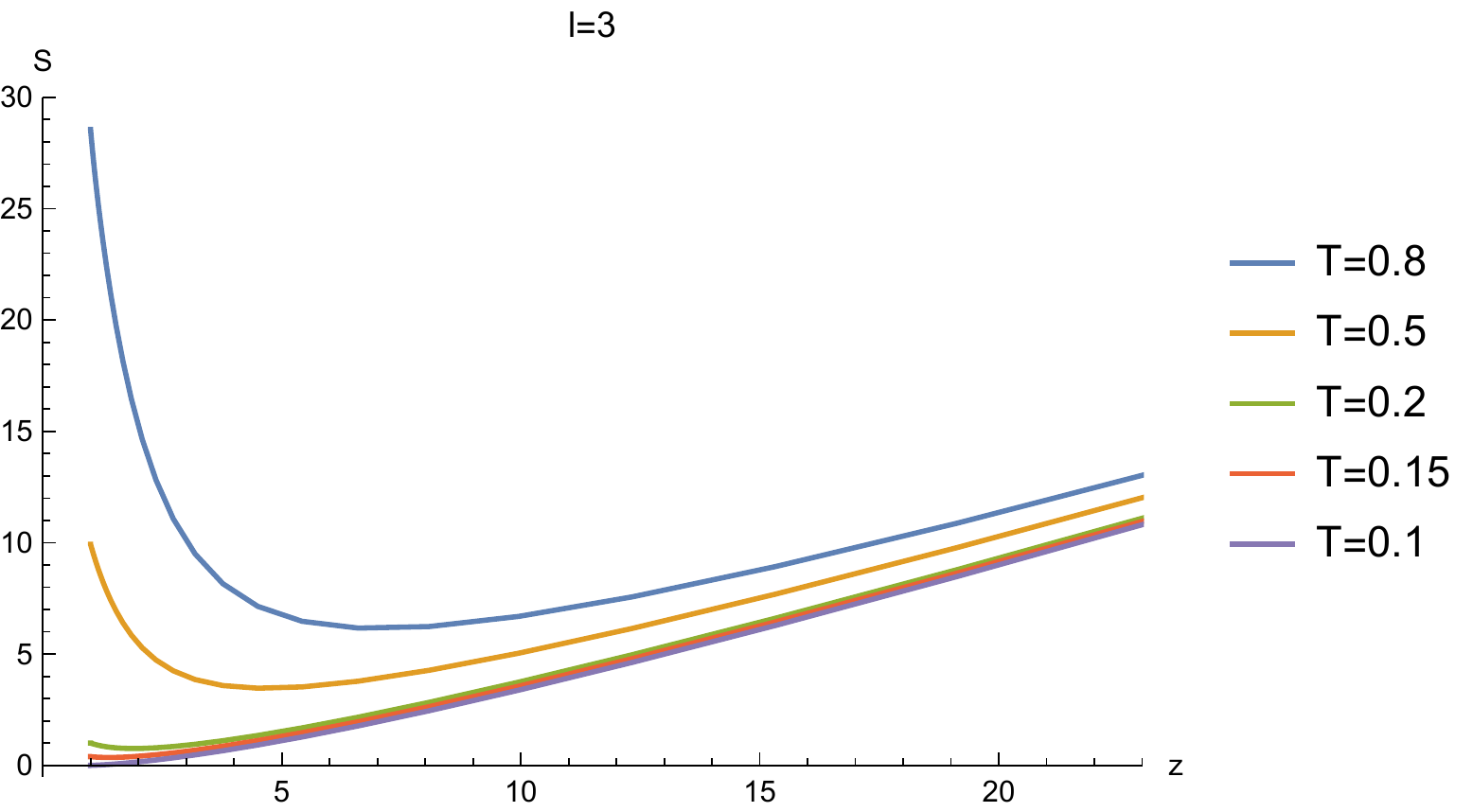}\ \\
    \caption{HEE vs $z$ for sample widths $l$ and temperatures $T$. (Top left) For $T=0.1$, HEE increases monotonically in the region of $l\geq1.8$. (Bottom left) For $T=1$, all curves of HEE shown here are non-monotonic. (Top right) For $l=2$, all curves of HEE shown here are non-monotonic. (Bottom right) For $l=3$, HEE increases monotonically in the region of $T< 0.15$.}
    \label{HEE vs z}
  }
\end{figure}

Also, from FIG.\ref{figHEE_differentz}, we see that some curves of HEE intersect each other (see the inserted plots). This observation indicates that for certain regions of $l$ and $T$, HEE is non-monotonic with $z$. To see this clearly, we plot HEE as a function of $z$ for sample widths $l$ and temperatures $T$ in FIG.\ref{HEE vs z}, of which we summarize the properties as follows.
%%%%
\begin{itemize}
  \item There is a region of large subsystem width and low temperature in which HEE increases monotonically with $z$. In this case, the degree of freedom with large $z$ is more entangled than that with small $z$. In particular, for $T=0.1$, as $z$ increases, HEE  increases monotonically in the region of $l\geq 1.8$, while for $l=3$, HEE increases monotonically with $z$ in the region of $T< 0.15$.
  \item When the subsystem width decreases or the temperature rises, the non-monotonic behavior of HEE with $z$ emerges. That is to say, as $z$ decreases, HEE decreases and arrives at a minimum value, and then goes up as $z$ further decreases.
\end{itemize}
So far, analytical understanding of the behaviors of HEE with $z$ is still absent. In addition, it is also desirable to test whether this behavior is universal in other Lifshitz gravity theories.
%%%p
%%%%%

%%%%
\subsection{Mutual Information}

In this subsection, we shall numerically study MI with symmetric and asymmetric configurations, since a more comprehensive configuration may provide more insight into the dual quantum system.

\subsubsection {Symmetric configuration}

%%%p
\begin{figure}[H]
  \center{
    \includegraphics[scale=0.45]{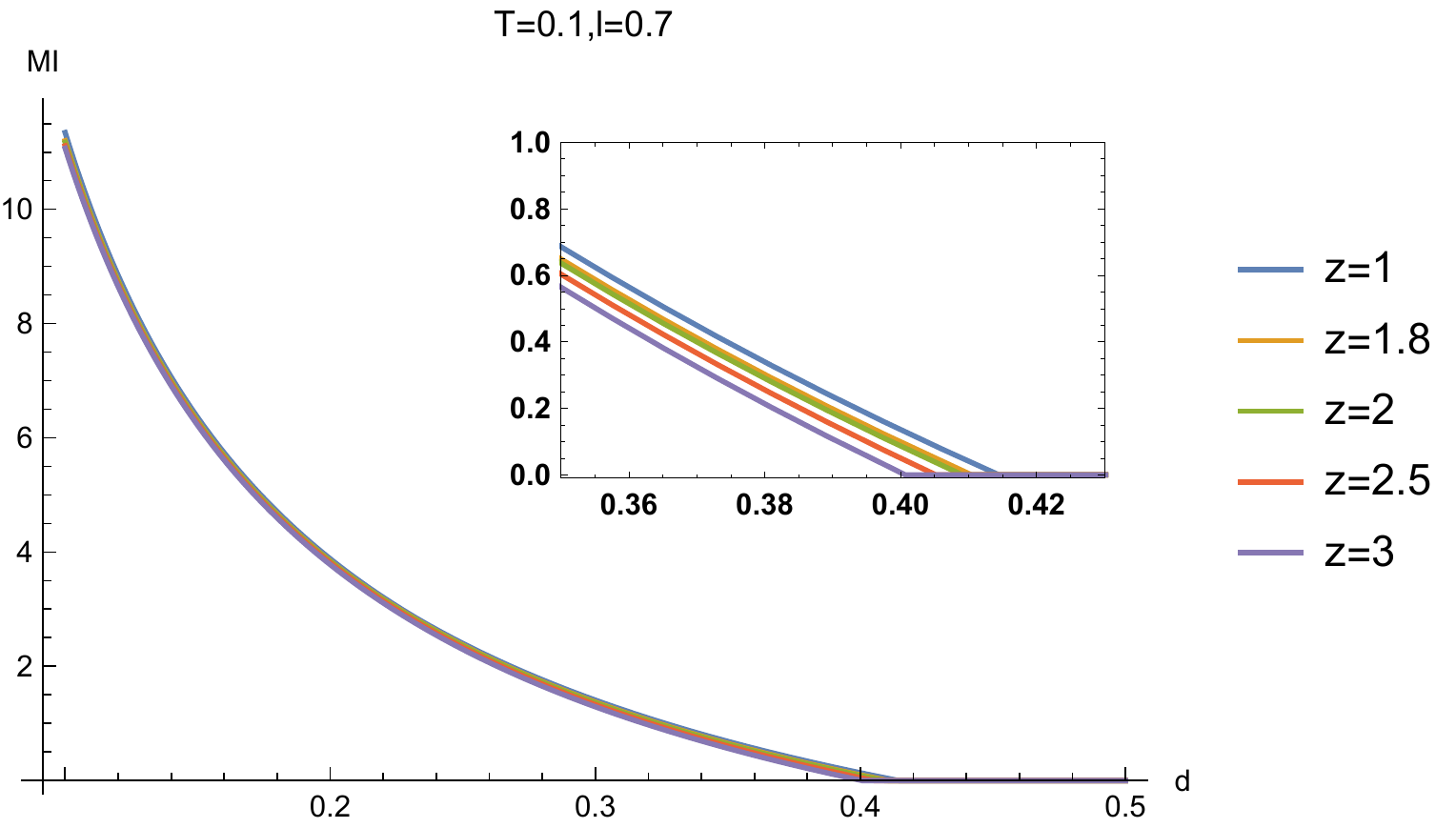}\ \hspace{0.5cm}
    \includegraphics[scale=0.45]{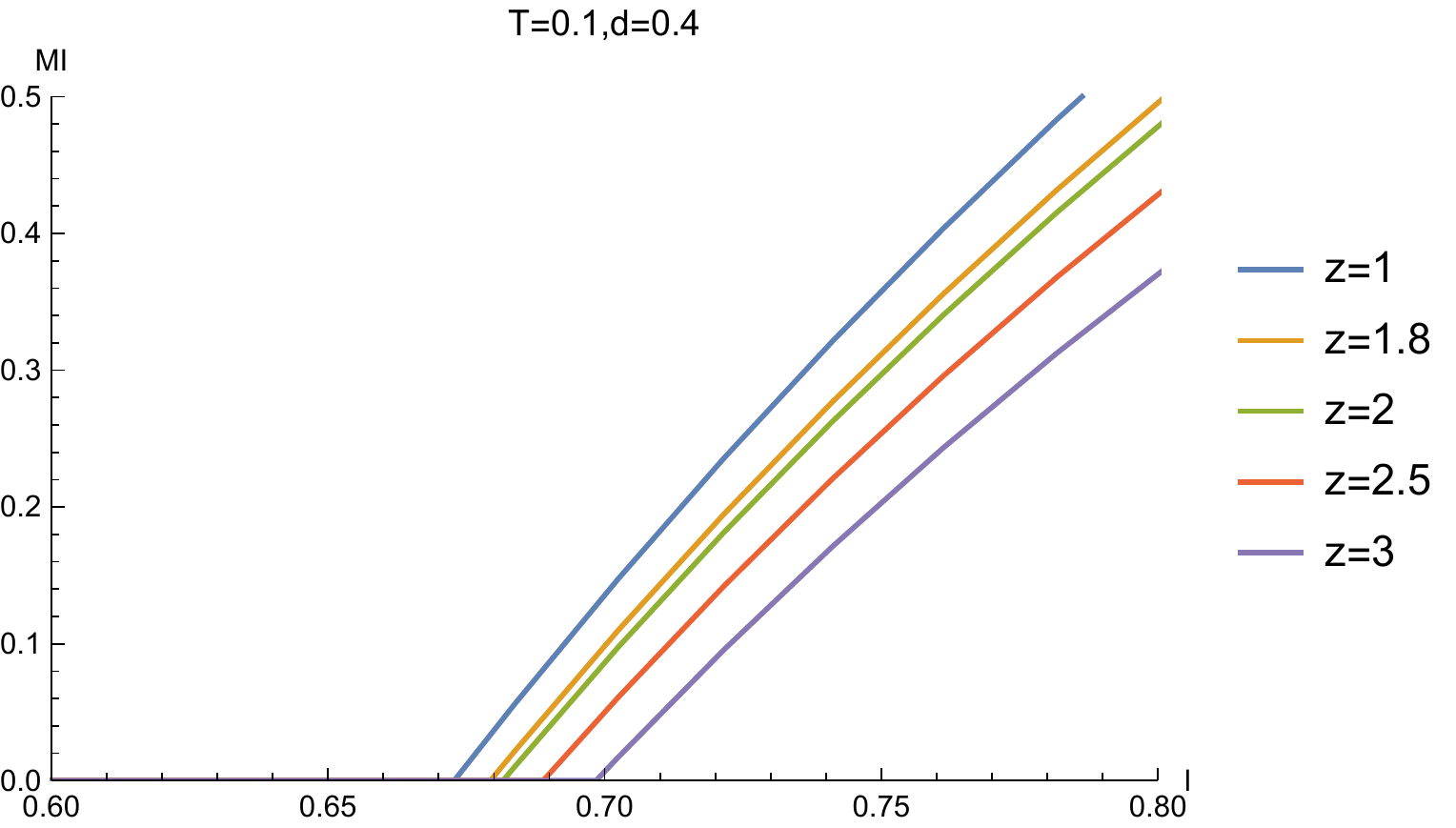}\
    \caption{MI vs separation scale $d$ with fixed system size $l$ (left) and vs $l$ with fixed $d$ (right) for different $z$. Here we have fixed the temperature $T=0.1$.}
    \label{symmetric MI vs ac or b}
  }
\end{figure}

We first study the case of symmetric configuration.
The left-hand plot in FIG.\ref{symmetric MI vs ac or b} demonstrates the behavior of MI with the separation scale $d$ for fixed system scale $l$ and different $z$ ($T=0.1$). We see that MI reduces as the separation scale grows.
When the separation scale grows further and goes beyond a certain critical value, MI decreases to zero.
This implies that a disentangling between the two sub-systems happens.
This is as expected since the correlation decays with the separation scale.
The behavior of MI with $l$ for fixed $d$ and different $z$ is also explored and shown in the right-hand plot in FIG.\ref{symmetric MI vs ac or b}. We can see that MI decreases with decreasing $l$, then reduces to zero as $l$ further decreases below a certain value. This behavior can be understood by the fact that shrinking the size of the subregion will reduce the degrees of freedom involved with the entanglement. These results from Lifshitz geometry are quantitatively consistent with those from RN-AdS geometry as well as from the Gubser-Rocha model \cite{Fu:2020oep}. Such a disentangling phase transition is a universal property of MI.

\begin{figure}[H]
  \center{
    \includegraphics[scale=0.35]{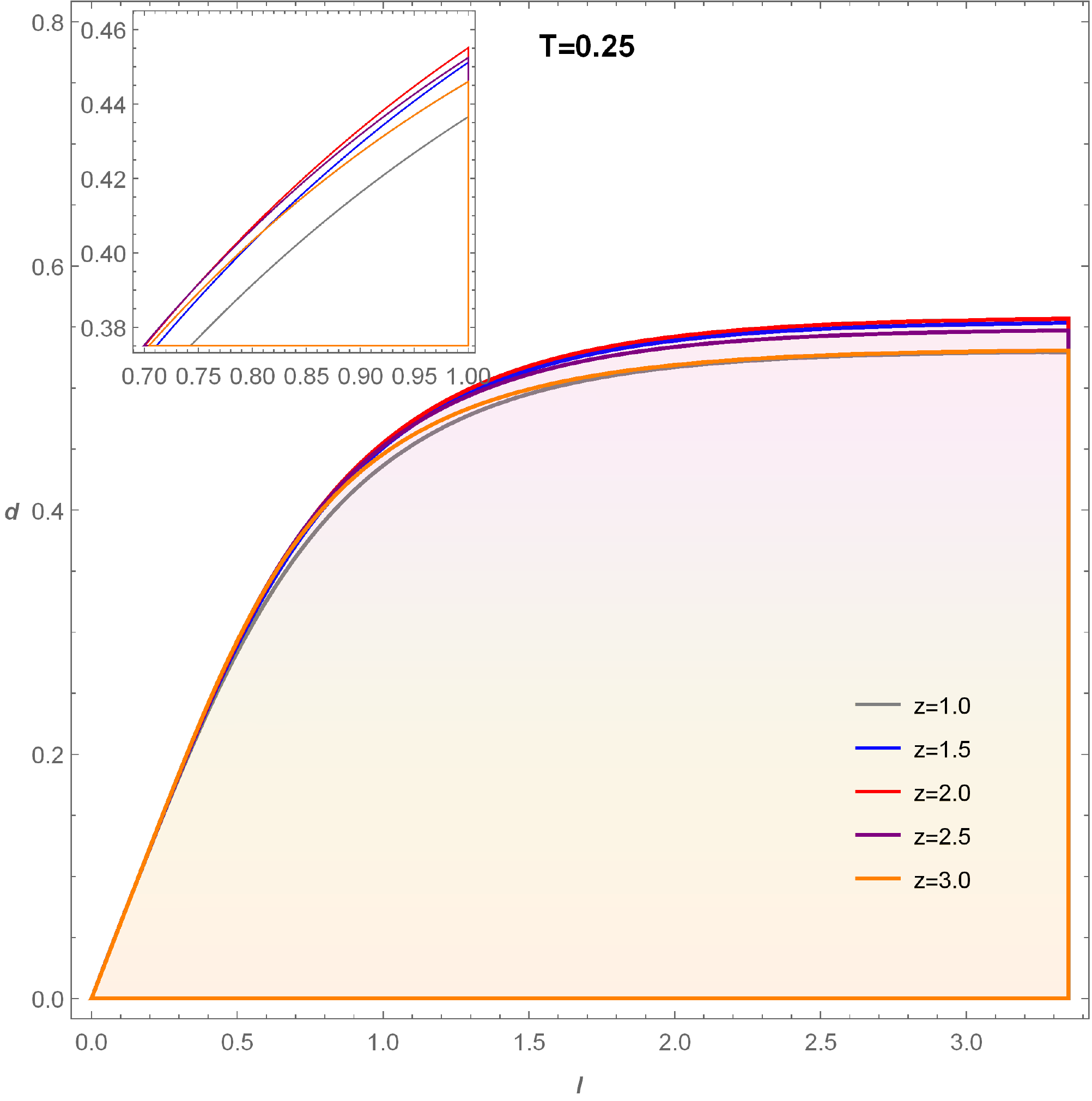}\ \\
    \caption{Parameter space $(l,d)$ for different $z$. The shaded region denotes non-zero MI.
    }
    \label{DvsL}
  }
\end{figure}

\begin{figure}[H]
  \center{
    \includegraphics[scale=0.55]{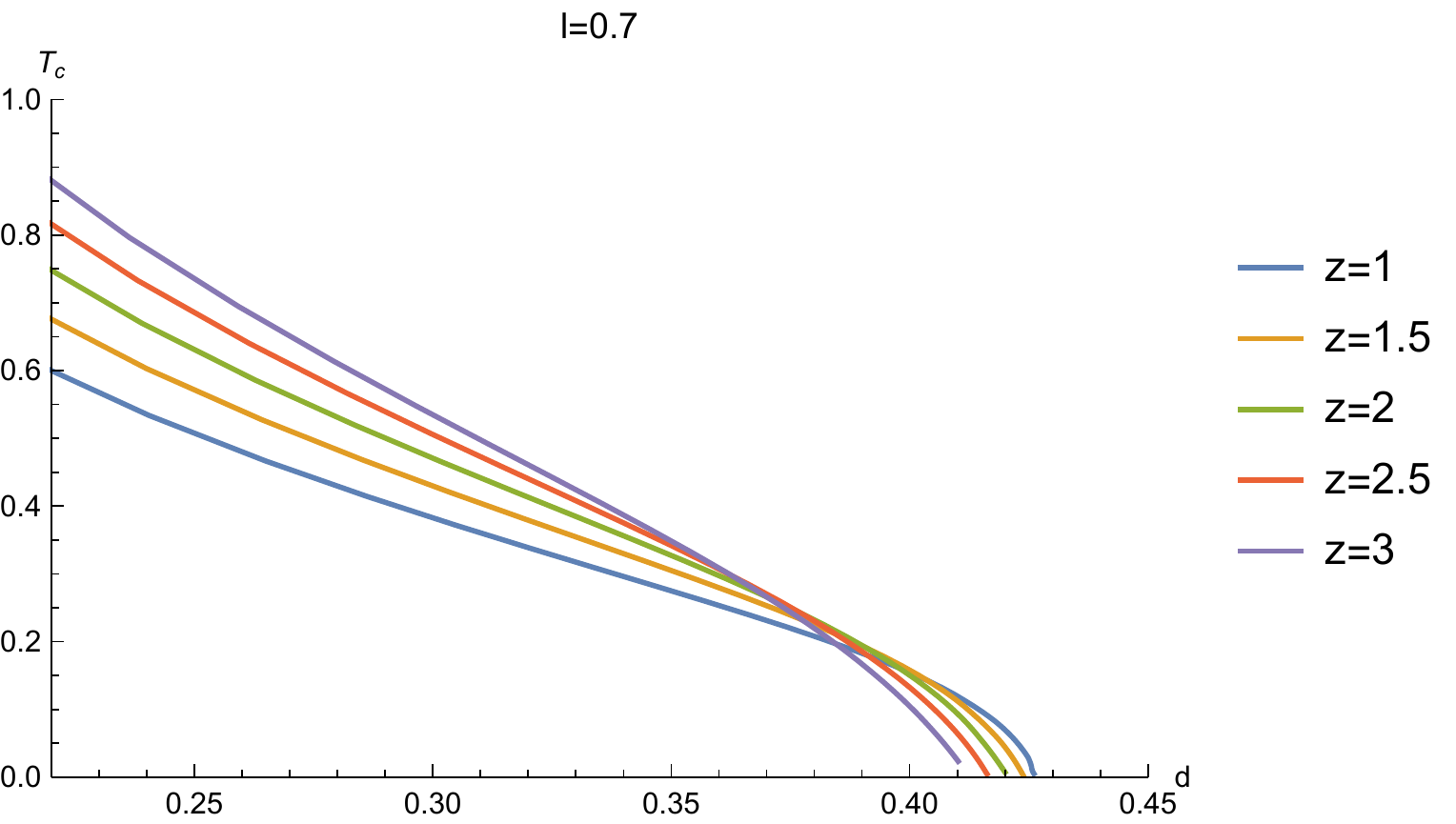}\
    \caption{Critical temperature $T_c$ vs $d$ for different $z$. Here $l$ has been fixed as $l=0.7$.}
    \label{sym_Tvsd_differentz}
  }
\end{figure}

Such a disentangling phase transition can be clearly demonstrated in the parameter space $(l,d)$ shown in FIG.\ref{DvsL}, in which non-vanishing MI is depicted by the shaded region. An obvious property is that for a fixed temperature, the critical lines for different $z$ approach a constant as $l$ grows. This means that the non-vanishing MI imposes a constraint on the separation scale $d$.
This result is in accordance with that of the Gubser-Rocha model studied in our previous work \cite{Fu:2020oep}. However, notice that these critical lines for different $z$ intersect each other (see also FIG.\ref{DvsL} for more obvious intersections). To confirm this observation, we also show the critical temperature $T_c$ as a function of $d$ for fixed $l$ and different $z$ (FIG.\ref{sym_Tvsd_differentz}). The curves for different $z$ do indeed intersect each other.
%%%%
%%%%%%%
%%%
%%%%%

\begin{figure}[H]
  \center{
    \includegraphics[scale=0.5]{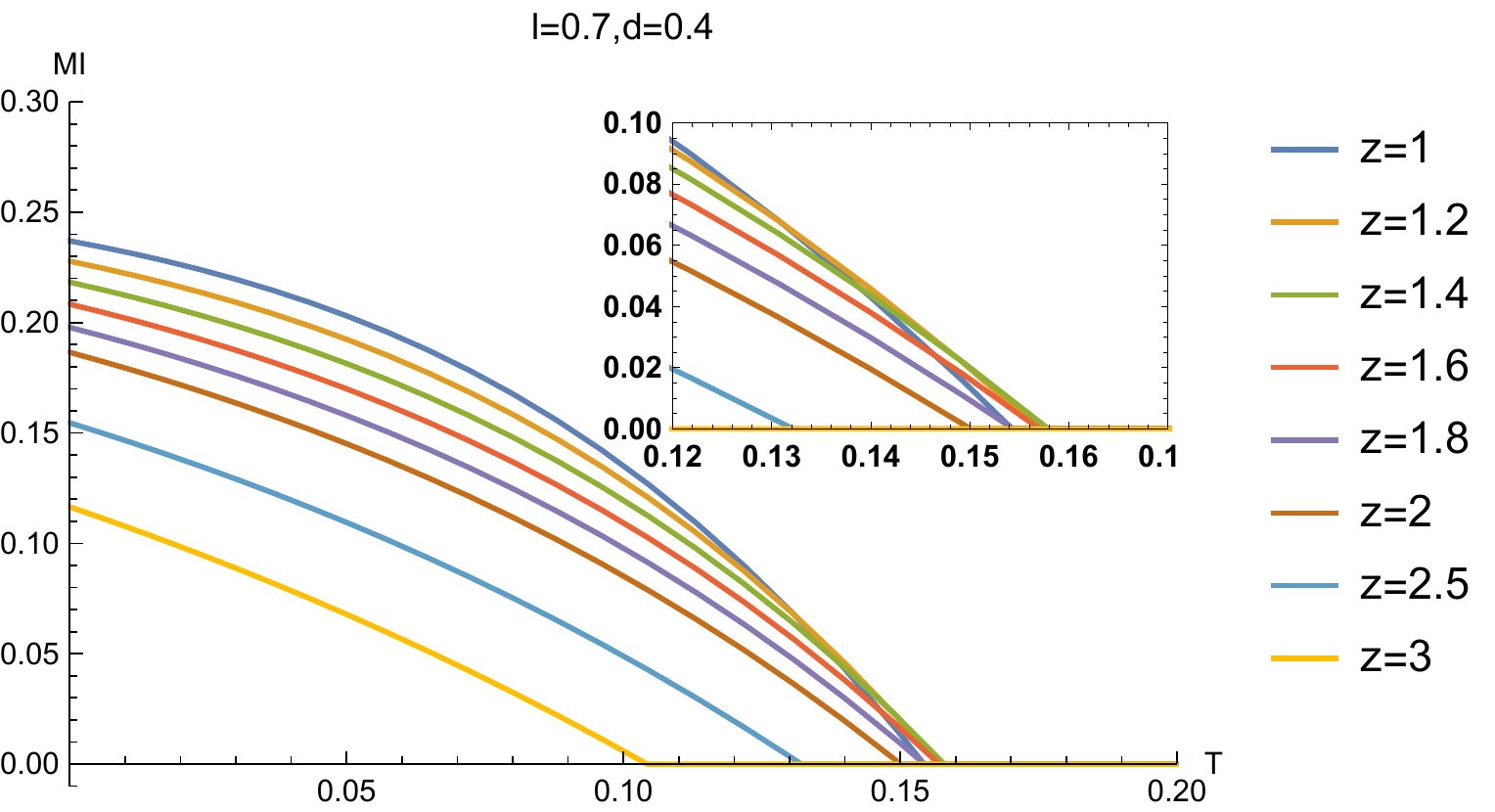}\ \hspace{0.5cm}
    \includegraphics[scale=0.5]{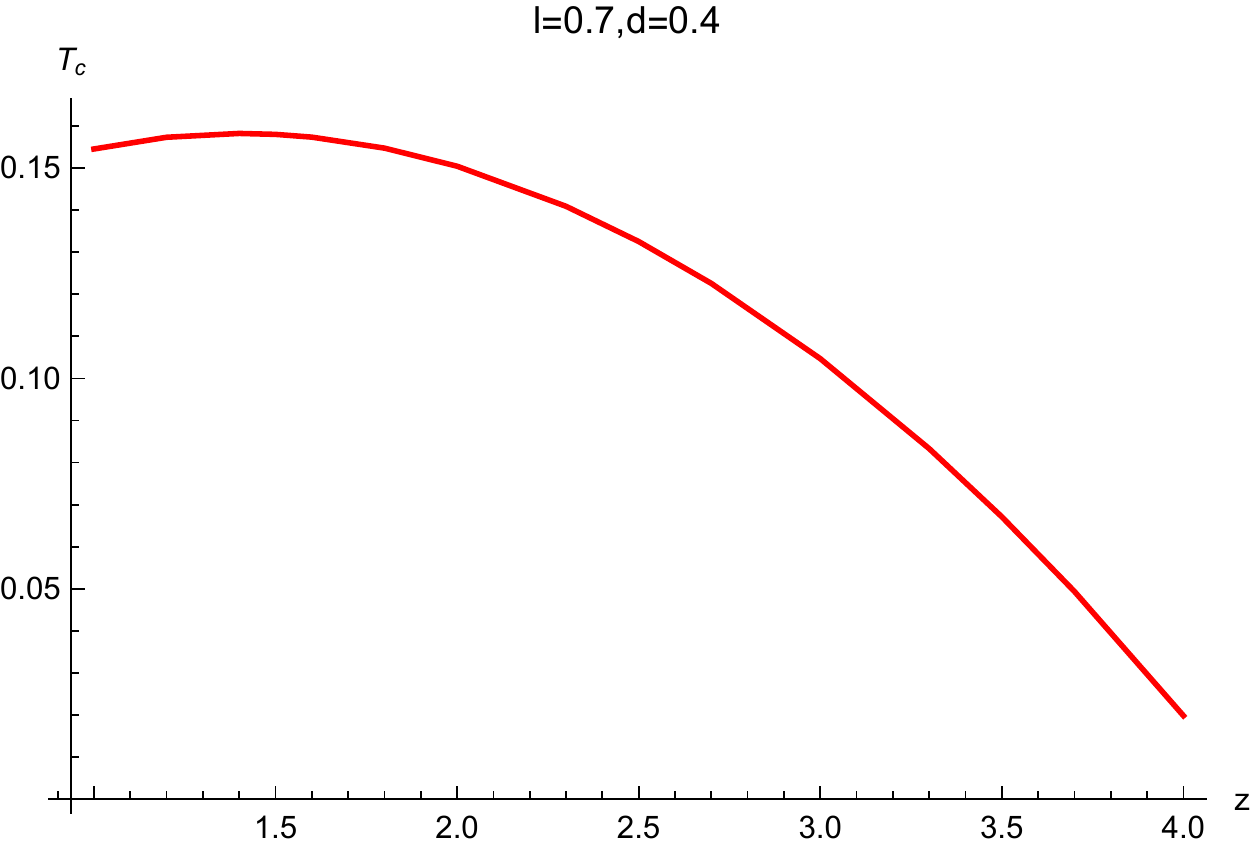}\\
    \caption{(left) MI vs $T$ for different $z$, and (right) the phase diagram $(z,T_c)$. The red line is the critical line, below which MI is non-zero and above which MI vanishes.
    }
    \label{symmetricMIvsT}
  }
\end{figure}

Now, we study the temperature behavior of MI. The Left-hand plot in FIG.\ref{symmetricMIvsT} shows the relation between MI and temperature for fixed configuration size at different $z$.
We see that when we heat up the system, MI reduces. Then as the temperature rises further and goes beyond certain critical value, MI reduces to zero. At this moment, a disentangling transition emerges.
This behavior is due to the fact that the thermal effects will destroy the quantum entanglement when the temperature is high.
This is a universal property which has been observed in previous works \cite{Fischler:2012uv,MolinaVilaplana:2011xt,Fu:2020oep}.
%%%%%
%%%%%p

In addition, we note that the curves of MI for some $z$ intersect each other near the disentangling critical temperature $T_c$ (left-hand plot in FIG.\ref{symmetricMIvsT}).
This indicates that the disentangling transition line does not change monotonically with $z$.
To confirm this observation, we show the phase diagram $(z,T_c)$ in the right-hand plot in FIG.\ref{symmetricMIvsT}.
Notice that as $z$ increases, the critical line first increases slightly, and then drops. We can conclude that as $z$ increases, the critical temperature of the disentangling phase transition becomes lower.
We can also infer that this non-monotonicity of MI with temperature is also inherited from that of HEE studied in previous subsection. This is reasonable because the MI is directly related to the HEE.
Notice moreover that in general, when $z$ is large, MI decreases with the increase of $z$. For small $z$, however, the opposite trend appears near the disentangling phase transition point.
%%%%
\begin{figure}[H]
  \center{
    \includegraphics[scale=0.5]{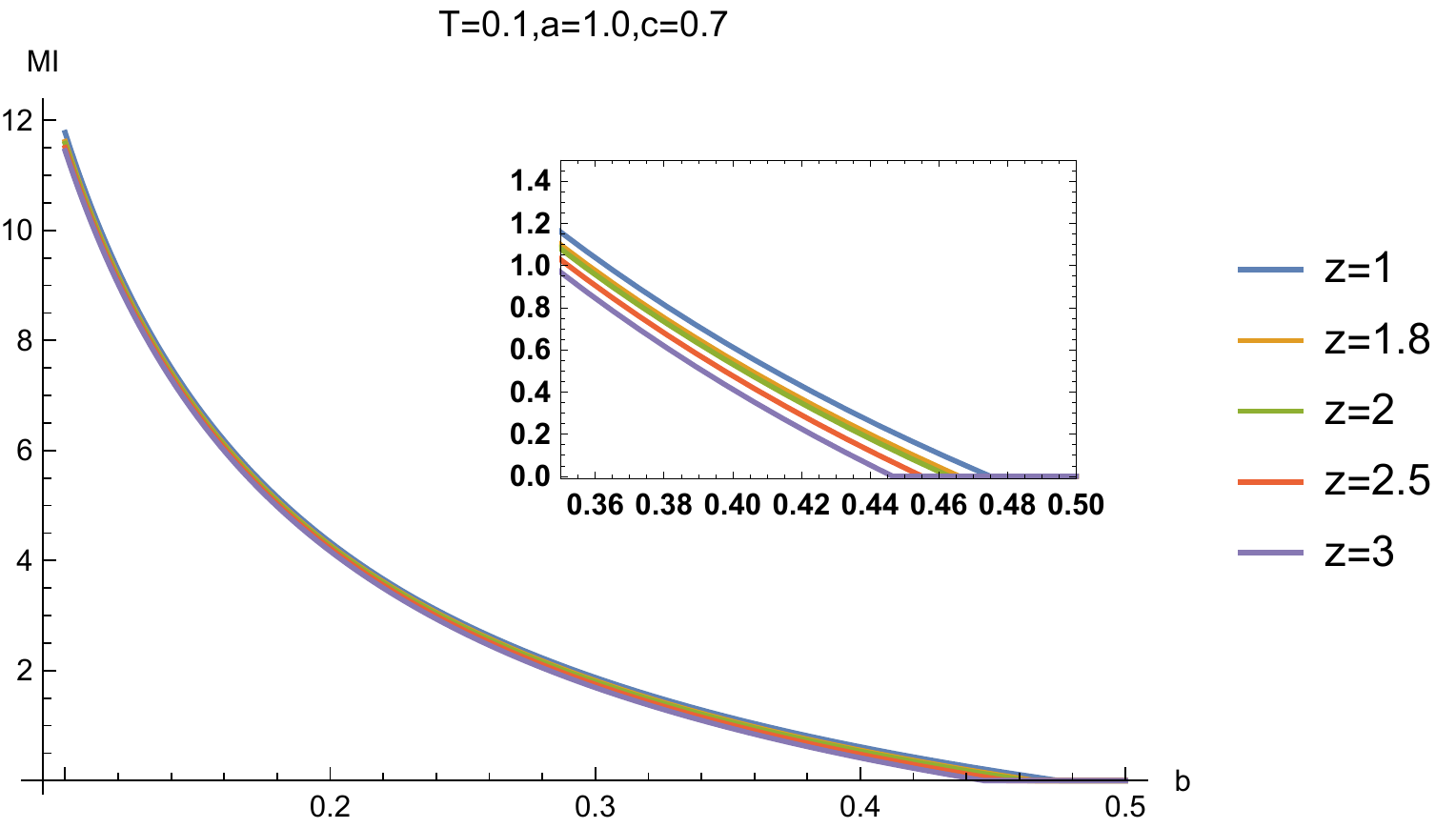}\ \hspace{0.5cm}
    \includegraphics[scale=0.5]{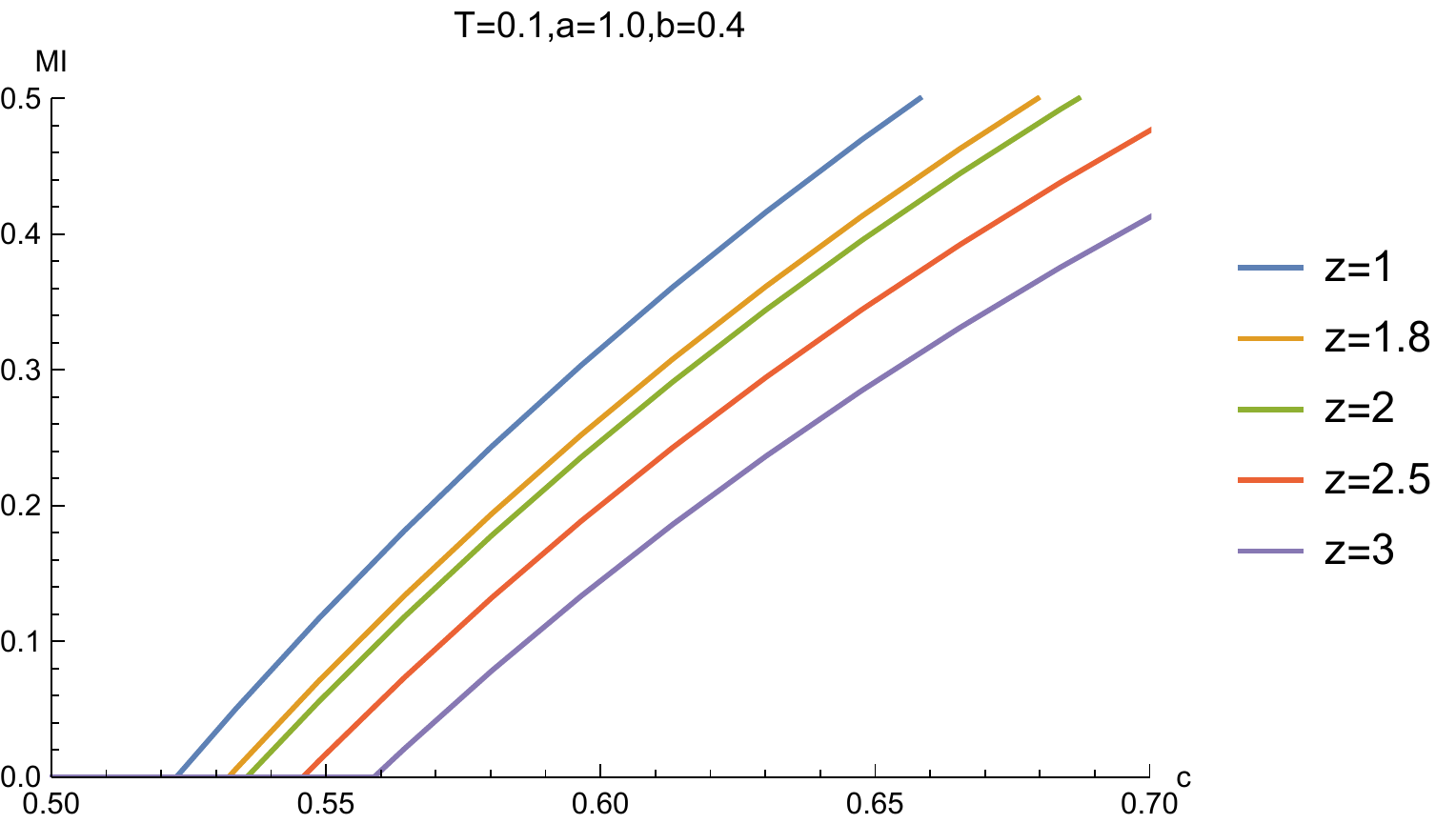}\ \\
    \caption{(left) MI vs separation scale $b$ with fixed system size $a=1.0$ and $c=0.7$ for different $z$, and (right) MI vs system size $c$ with $a=1.0$ and $b=0.4$ for different $z$.
      Here the temperature is fixed as $T=0.1$.}
    \label{asymmetric MI vs ac or b}
  }
\end{figure}
%%%%%p

\subsubsection {Asymmetric configuration}

The schematic configuration for computing MI can be seen in FIG.\ref{EoPcartoonpicture}, for which the sizes of two subsystems are unequal, i.e., $a\neq c$. We show the results for MI in FIG.\ref{asymmetric MI vs ac or b}, showing that a disentangling phase transition happens as the separation scale increases or the subsystem size decreases.
This property is universal, which is independent of the configuration.
%%%%
\begin{figure}[H]
  \centering
  \includegraphics[width=8cm]{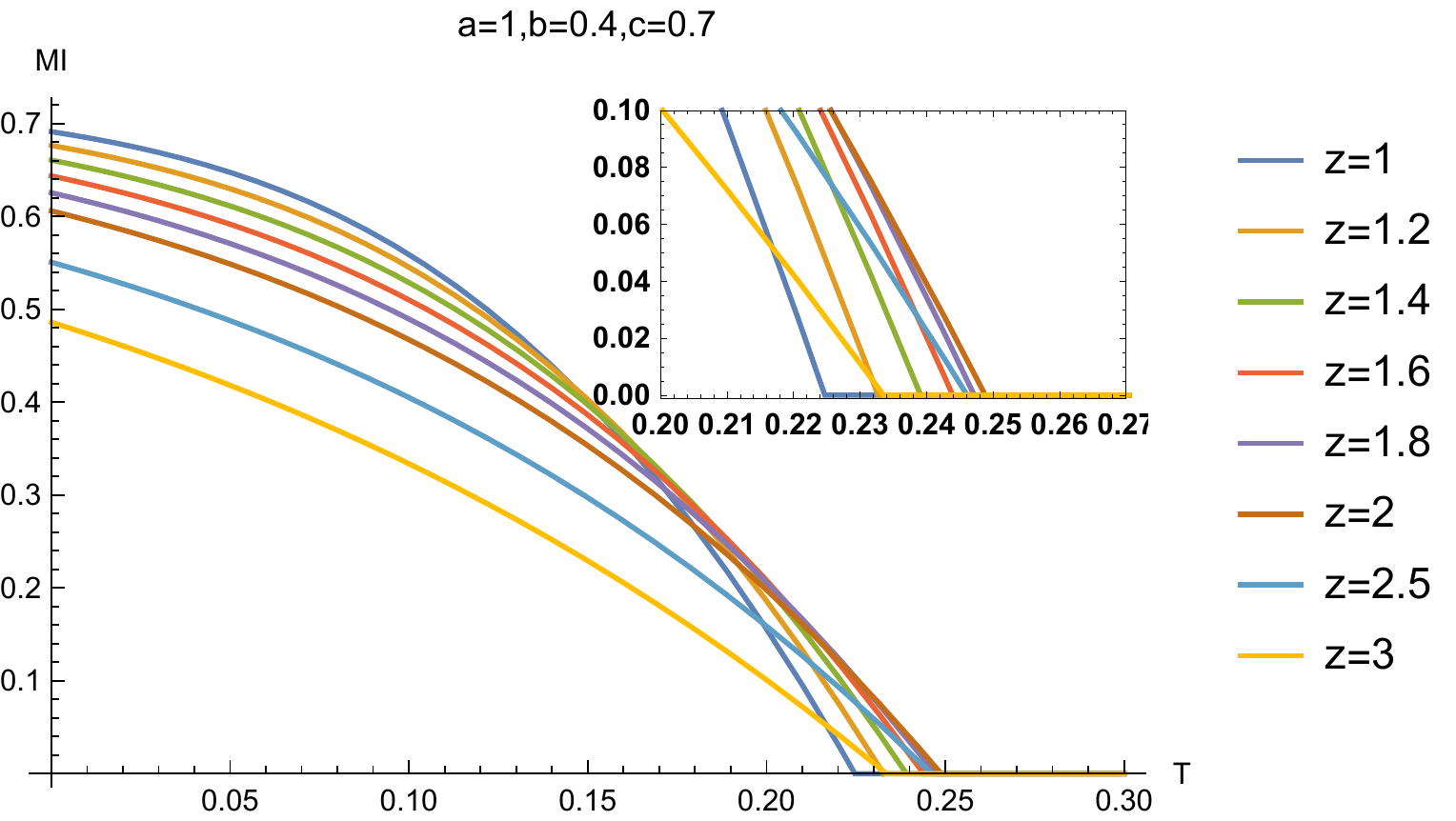}\ \hspace{0.5cm}
  \includegraphics[width=7cm]{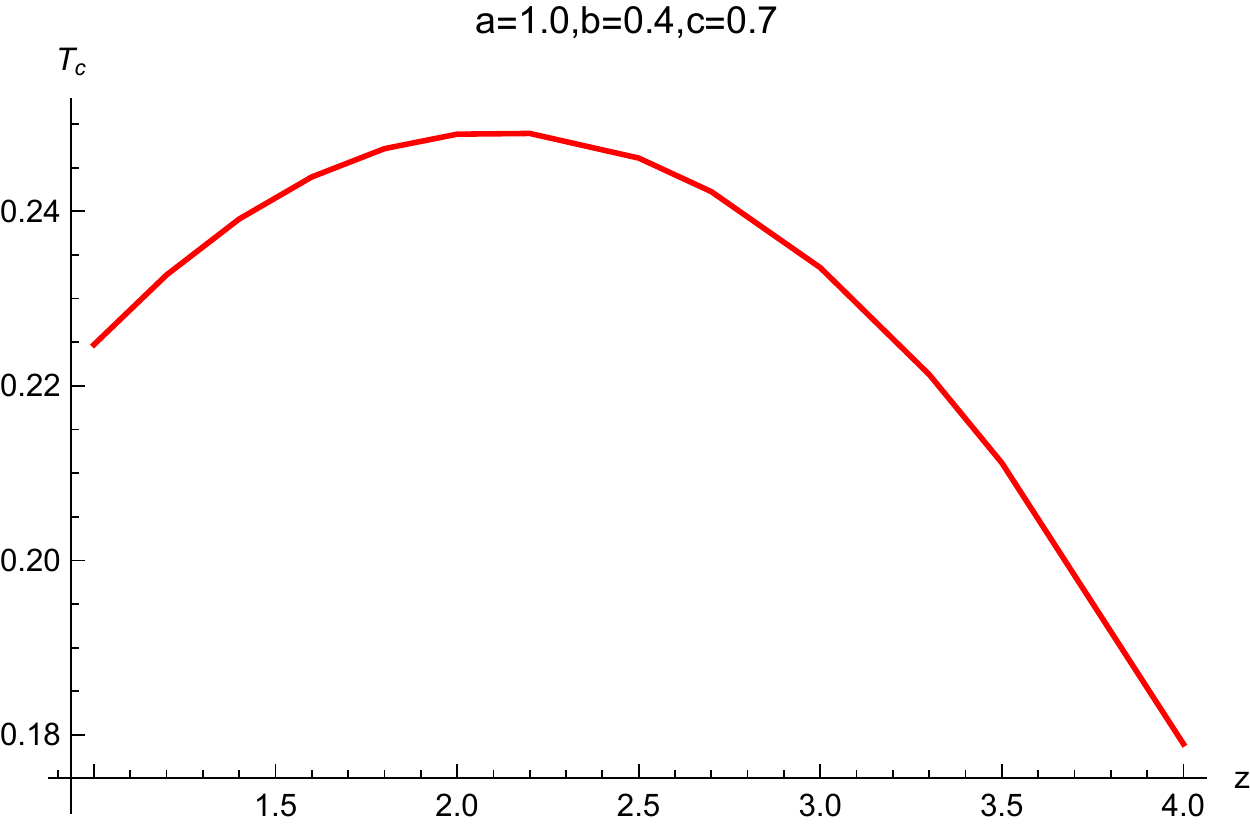}\ \\
  \caption{
    (left) MI vs $T$ for different $z$, and (right) the phase diagram $(z,T_c)$. The red line is the critical line, below which MI is non-zero and above which MI vanishes.}
  \label{asymmetric MI vs T}
\end{figure}
%%%%%

We then study the relation between MI and temperature for fixed configuration size at different $z$
(left-hand plot in FIG.\ref{asymmetric MI vs T}).
Similar to the case with symmetric configuration, we see that as we heat up the holographic system,
a disentangling phase transition happens. This indicates that such a disentangling phase transition is also independent of the configuration.

Also, the curves of MI for some $z$ intersect each other near the critical point of disentangling phase transition, which is also similar to the case of symmetric configuration.
Therefore we conclude that the critical points of a holographic Lifshitz system do not show a monotonic relationship with $z$.
Further, we plot the phase diagram $(z,T_c)$ in the right-hand plot in FIG.\ref{asymmetric MI vs T}.
Compared with the symmetric configuration, before the critical line drops with the increase of $z$, there is a larger region of $z$ in which the critical line rises as $z$ increases. Therefore, the non-monotonic region of $z$ depends on the configuration. This is reasonable because, as observed in the subsection above, the non-monotonic behavior of HEE also depends on the system width $l$ and the temperature $T$.

\begin{figure}[H]
  \centering
  \includegraphics[width=8cm]{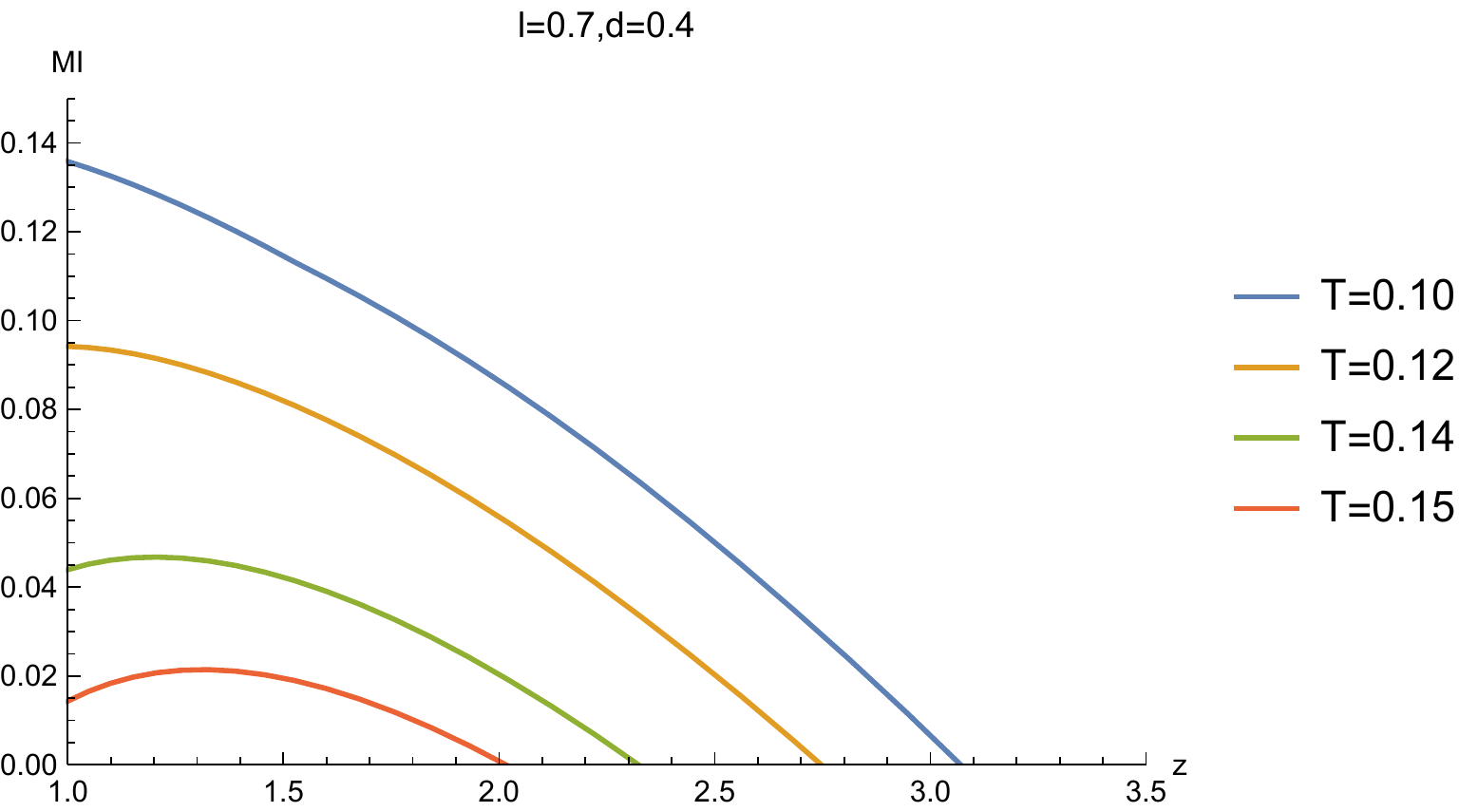}\ \hspace{0.2cm}
  \includegraphics[width=8cm]{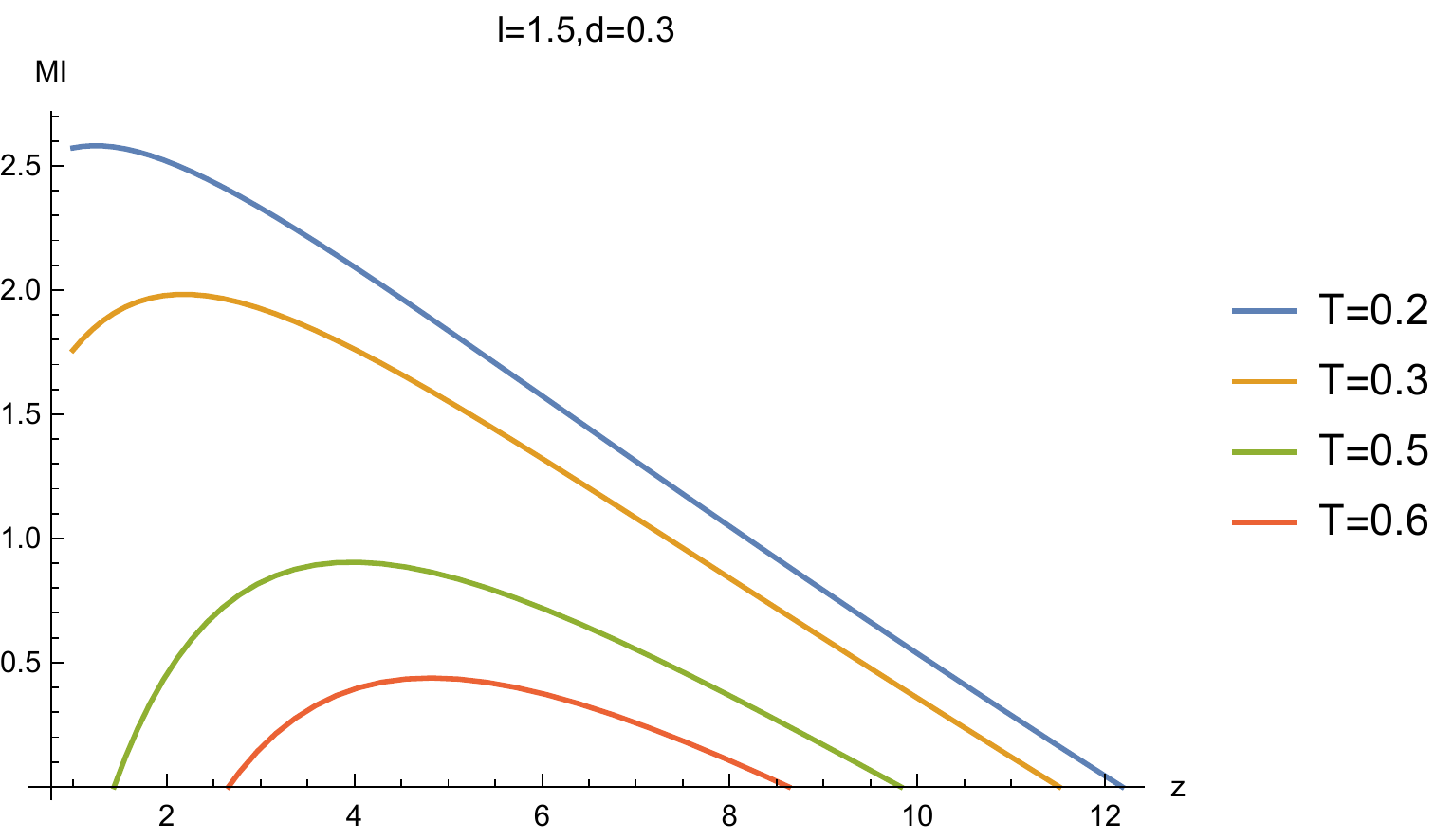}\ \\
  \caption{MI as a function of $z$ for selected configuration parameters ($l$ and $d$) and temperatures.}
  \label{MIvsz}
\end{figure}

In the right-hand plots of FIG.\ref{symmetricMIvsT} and FIG.\ref{asymmetric MI vs T}, the critical line of the disentangling phase transition exhibits non-monotonic behavior. We expect that MI as a function of $z$ also exhibits non-monotonic behavior for some system configuration parameters and temperatures. To this end, we plot MI as a function of $z$ in FIG.\ref{MIvsz}. In the left-hand plot, with fixed $l=0.7$ and $d=0.4$, we find that when the temperature is low, MI decreases monotonically with the increase of $z$ and vanishes when $z$ is beyond some critical temperature, which means that a disentangling phase transition happens. As the temperature rises, the non-monotonic behavior emerges. That is to say, as $z$ increases, MI first increase and then gradually decreases to zero and the disentangling phase transition happens. For another configuration ($l=1.5$ and $d=0.3$, see the right-hand plot in FIG.\ref{MIvsz}), we observe the novel phenomenon that a dome-shaped diagram emerges when the temperature is high. This means that when $z$ is smaller than some critical value or larger than some critical value, the biparty subsystem is disentangled in terms of MI.
%%%%p
%%%%%

%%%%
\subsection{Holographic Entanglement of Purification}

\begin{figure}[H]
  \center{
    \includegraphics[scale=0.5]{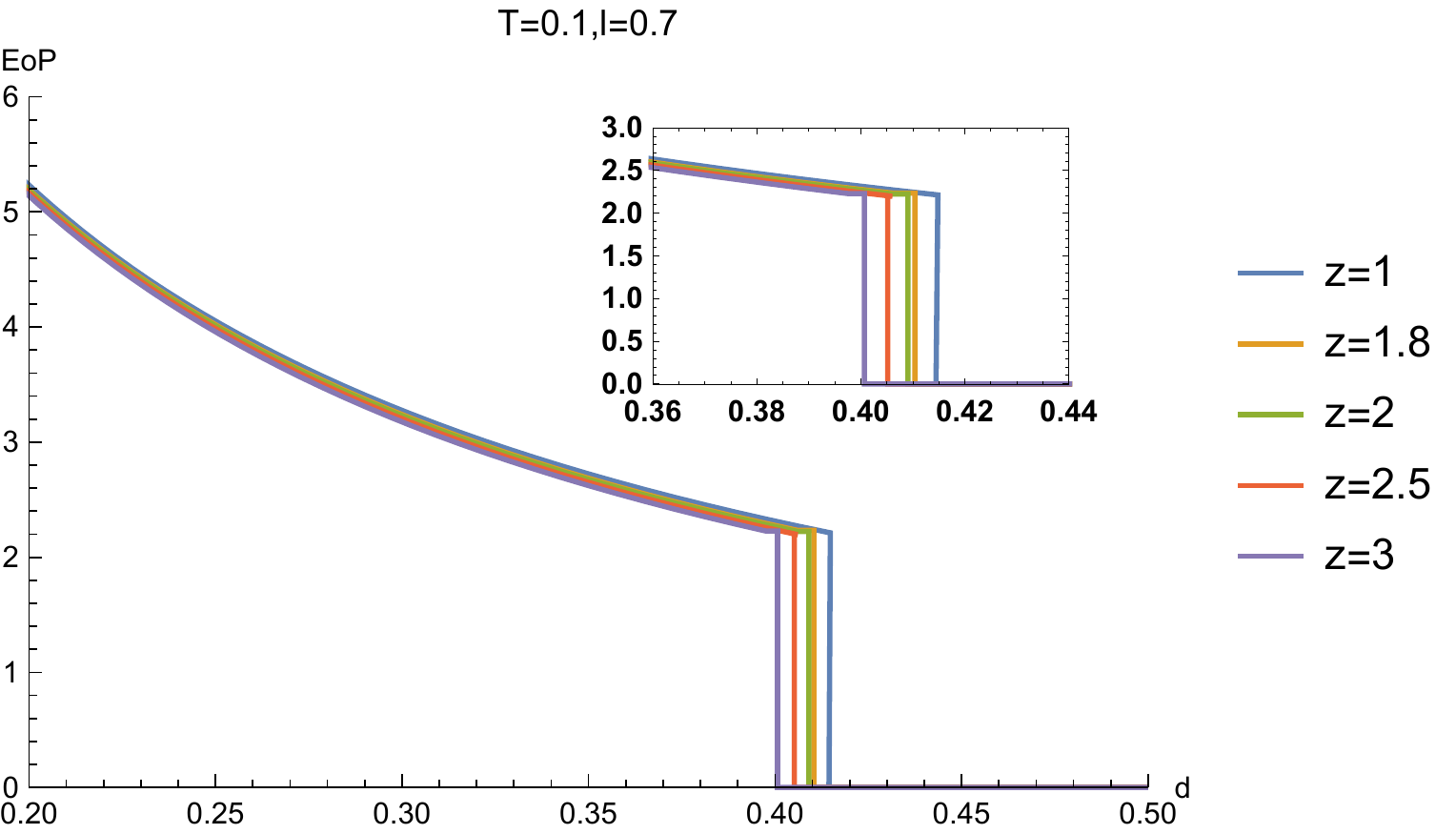}\ \hspace{0.5cm}
    \includegraphics[scale=0.5]{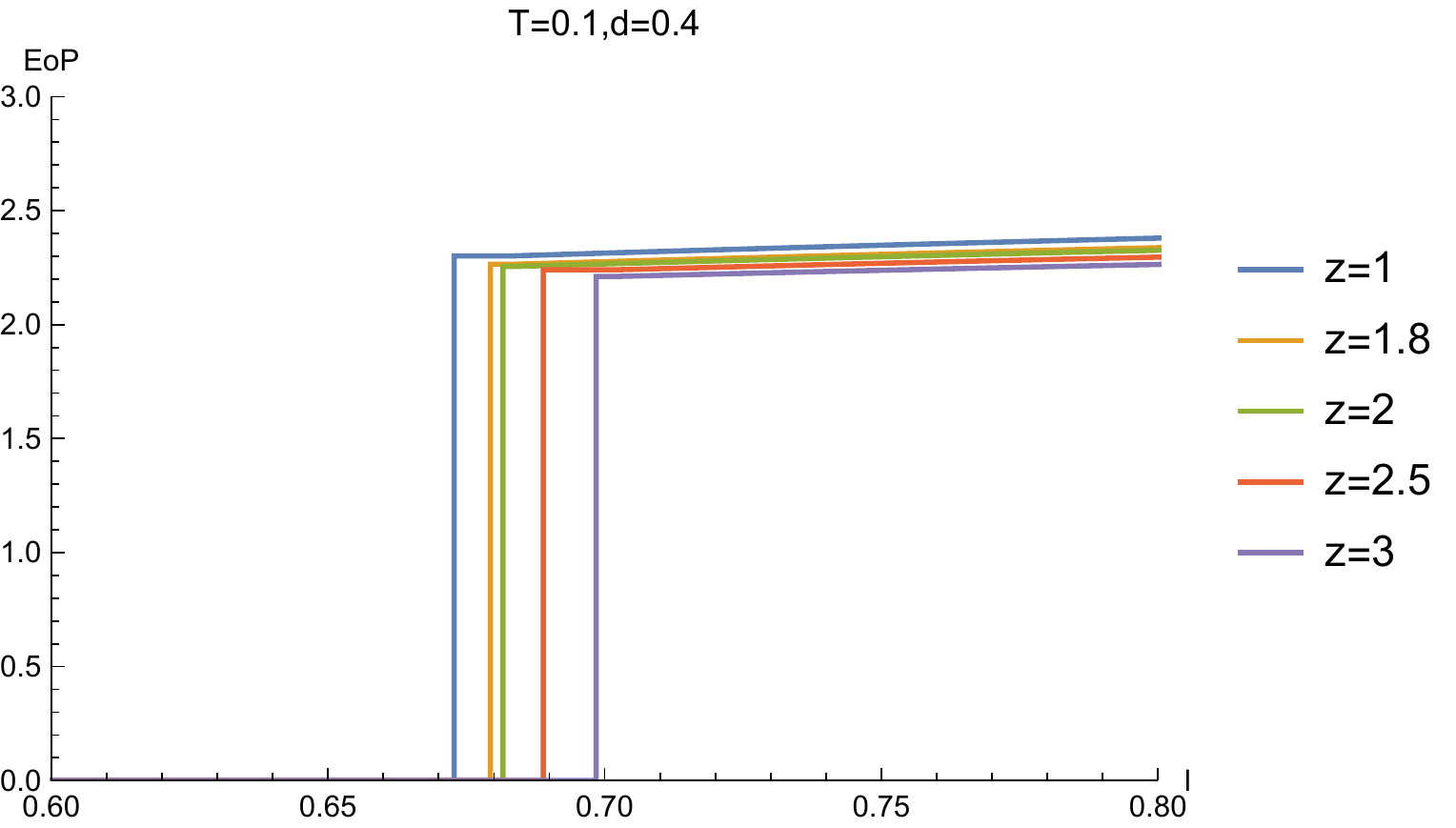}\ \\
    \caption{(left) $EoP$ vs separation scale $d$ for different $z$ ($l=0.7$), and
      (right) $EoP$ vs system size $l$ for different $z$ ($d=0.4$ ).
      Here, we have set $T=0.1$.}
    \label{symmetric EoP vs ac or b}
  }
\end{figure}

For a symmetric configuration, the calculation of EoP is simply the area of the vertical line linking the tops of the minimum surfaces (see left-hand plot in FIG.\ref{EoPcartoonpicture}).
In FIG.\ref{symmetric EoP vs ac or b}, we show EoP in a holographic Lifshitz system as a function of $d$ (left-hand plot)
and $l$ (right-hand plot).
As we have seen in the above section, when the two subsystems $A$ and $B$ are far away from each other, MI reduces to zero.
As a result, the entanglement wedge $\Gamma_{AB}$ is disconnected and hence $E_p=0$.
As the two subsystems approach each other, MI obtains a finite positive value and the entanglement wedge becomes connected.
Correspondingly, the EoP suddenly increases to a finite value.
After that, EoP rises further as both subsystems approach more closely.
%%%
%%%%%p

If we fix the separation scale $d$ of the subsystems, we find that EoP becomes zero when the subsystem is small.
This is also because in this region, MI is zero and the entanglement wedge $\Gamma_{AB}$ becomes disconnected.
As the size of the subsystem grows, the EoP suddenly increases to a finite value, for which an entangling phase transition emerges. Then, as the system size increases further, the EoP grows slowly.

\begin{figure}[H]
  \centering
  \includegraphics[width=10cm]{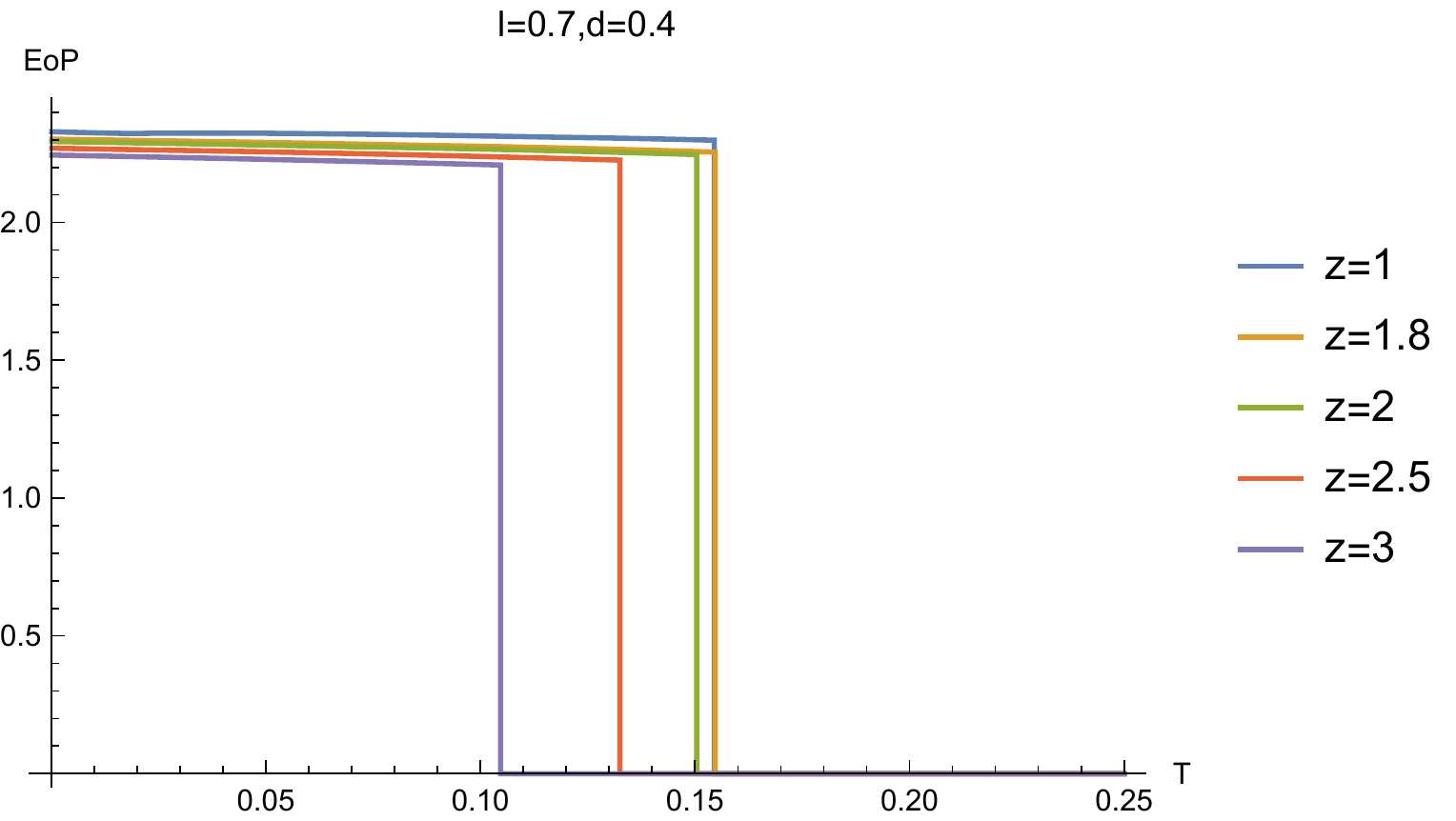}
  \caption{EoP $E_p$ vs $T$ for different $z$ ($l=0.7$ and $d=0.4$).}%%%%%
  \label{symmetric EoP vs T}
\end{figure}

We then study the temperature behavior of EoP, which is shown in FIG.\ref{symmetric EoP vs T}.
We see that in the high temperature region, EoP vanishes.
The reason is the same as  discussed above; MI is also zero in this region and so the entanglement wedge $\Gamma_{AB}$ disconnects. From the dual view, this is because increasing the temperature can destroy the entanglement between two separate sub-regions.
As the temperature drops beyond a critical point, EoP suddenly increases to a finite value.
As we further cool down the system, EoP grows slowly with the decrease of temperature.

%%%%%%%%%%%%%%
%%%%%%p
\begin{figure}[H]
  \centering
  \includegraphics[width=8cm]{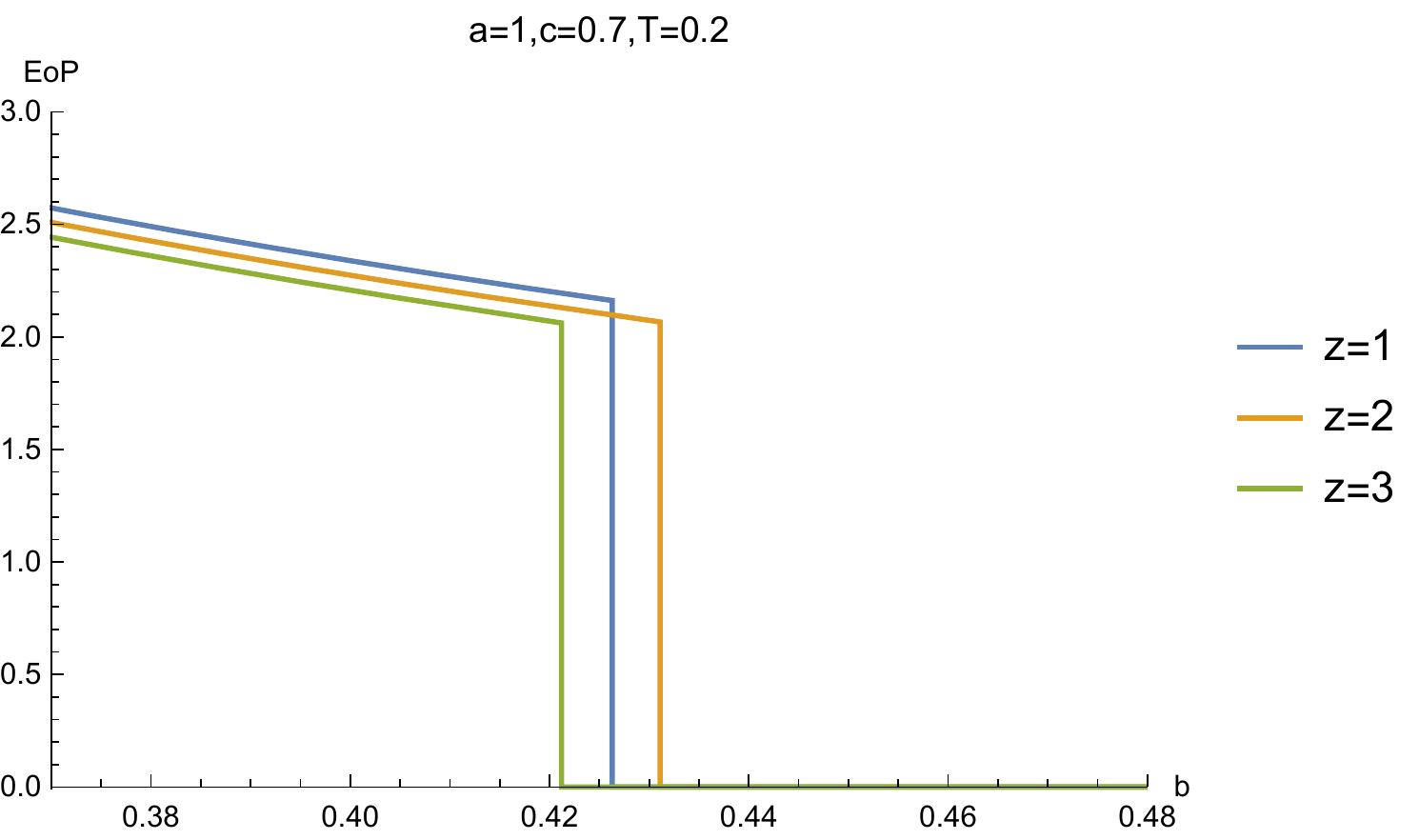}\ \hspace{0.2cm}
  \includegraphics[width=8cm]{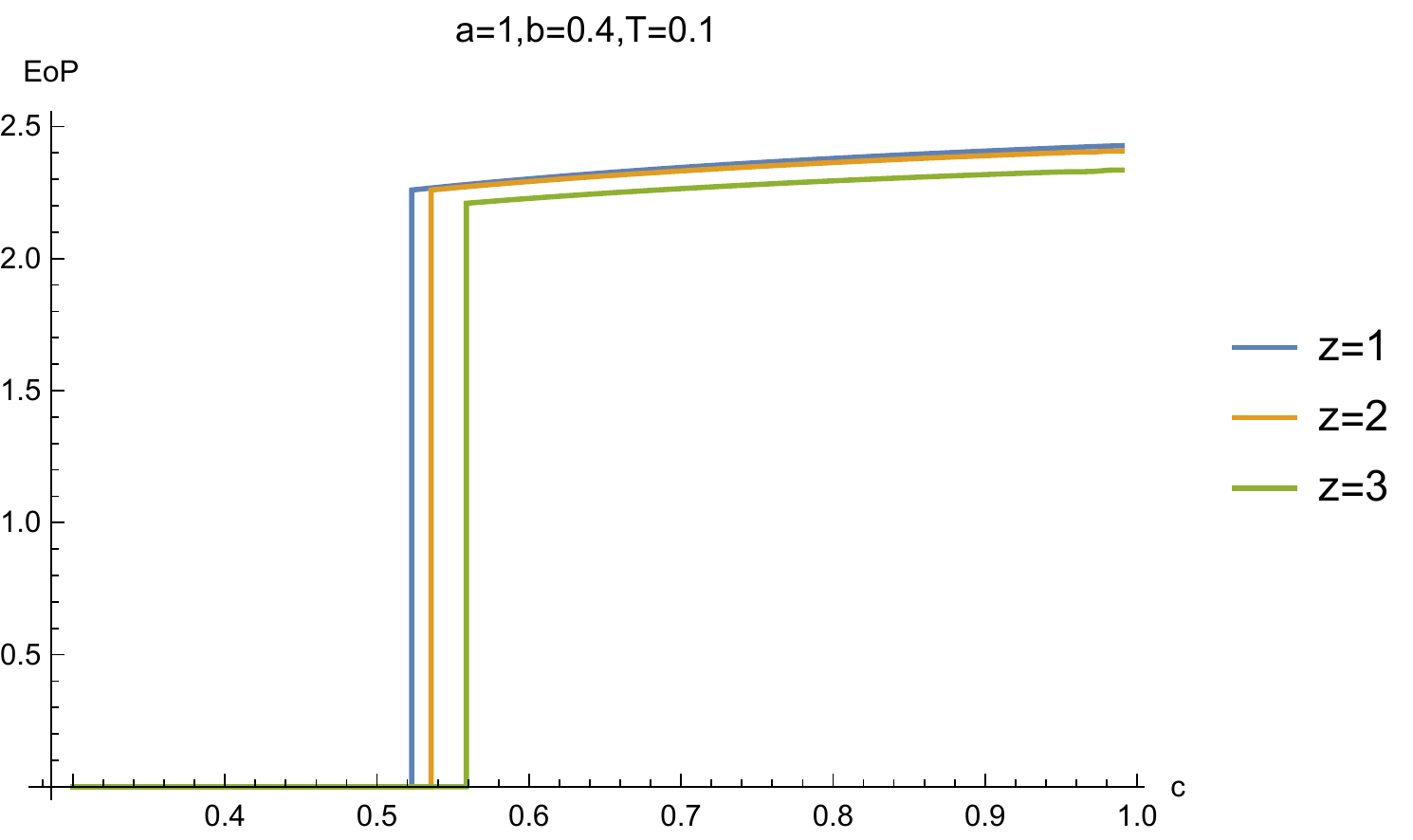}\ \\
  \caption{(left) $EoP$ vs $b$ for different $z$ ($a=1$, $c=0.7$ and $T=0.1$ ), (right) $EoP$ vs $c$ for different $z$ ($a=1$, $b=0.4$ and $T=0.1$).}
  \label{asym EoP vs b and c}
\end{figure}
%%%%%%

\begin{figure}[H]
  \centering
  \includegraphics[width=8cm]{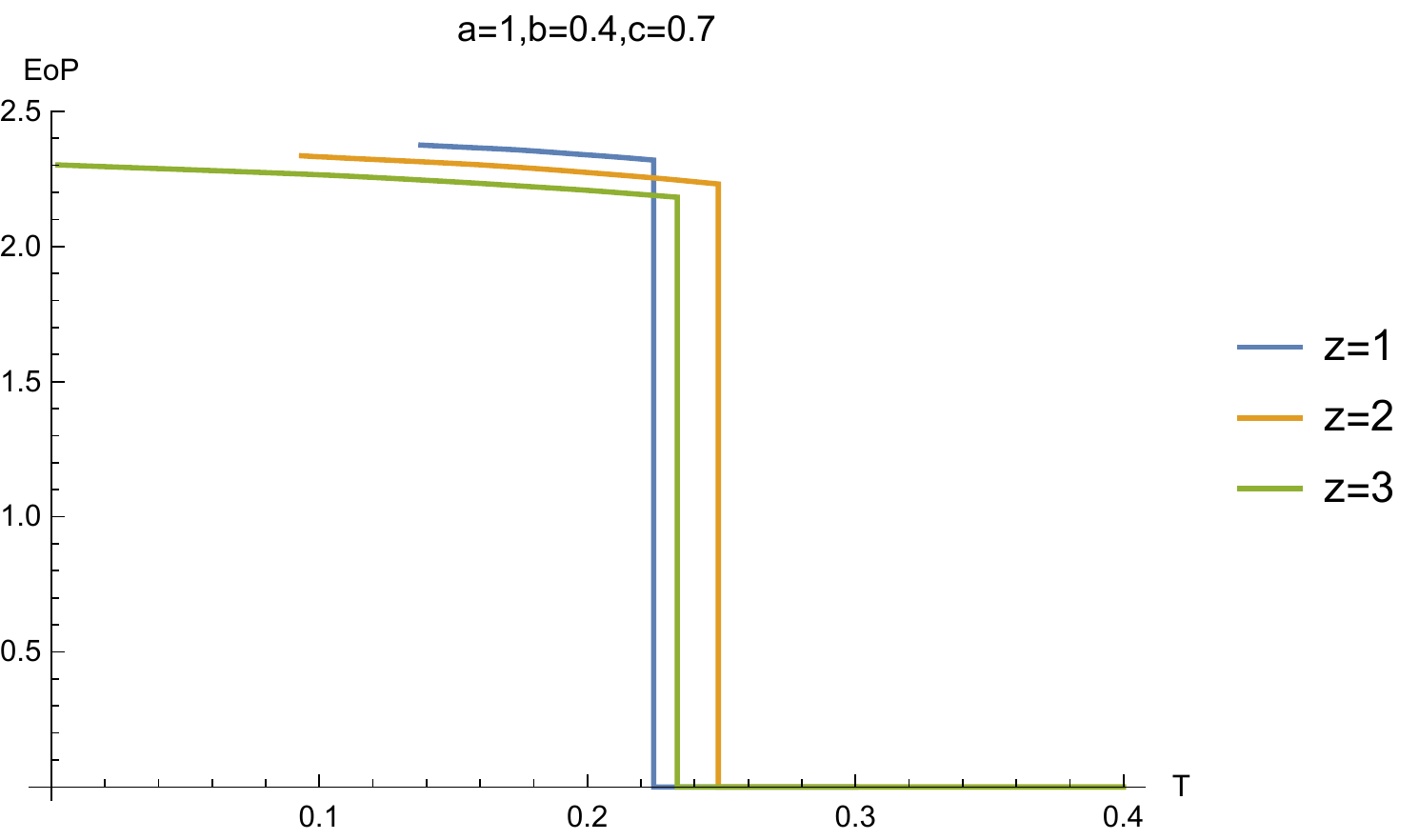}
  \caption{$EoP$ vs $T$ for different $z$ ($a=1.0$,$b=0.4$,$c=0.7$).}
  \label{asym EoP vs T}
\end{figure}

The behaviors of EoP for the asymmetric configuration are very similar to the symmetric configuration.
That is to say, EoP decreases with the increase of system size and suddenly decreases to zero, which indicates it enters into a disentangling state (left-hand plot in FIG.\ref{asym EoP vs b and c}).
In contrast to this process, EoP is zero when the system size is small.
When the system size increases to a critical value, EoP enters into an entangling state
and then EoP grows slowly as the system size increases (right-hand plot in FIG.\ref{asym EoP vs b and c}).
In addition, as the temperature drops, EoP slowly decreases and suddenly decreases to zero when the temperature reaches a critical value (FIG.\ref{asym EoP vs T}).

%%%%
\begin{figure}[H]
  \centering
  \includegraphics[width=8cm]{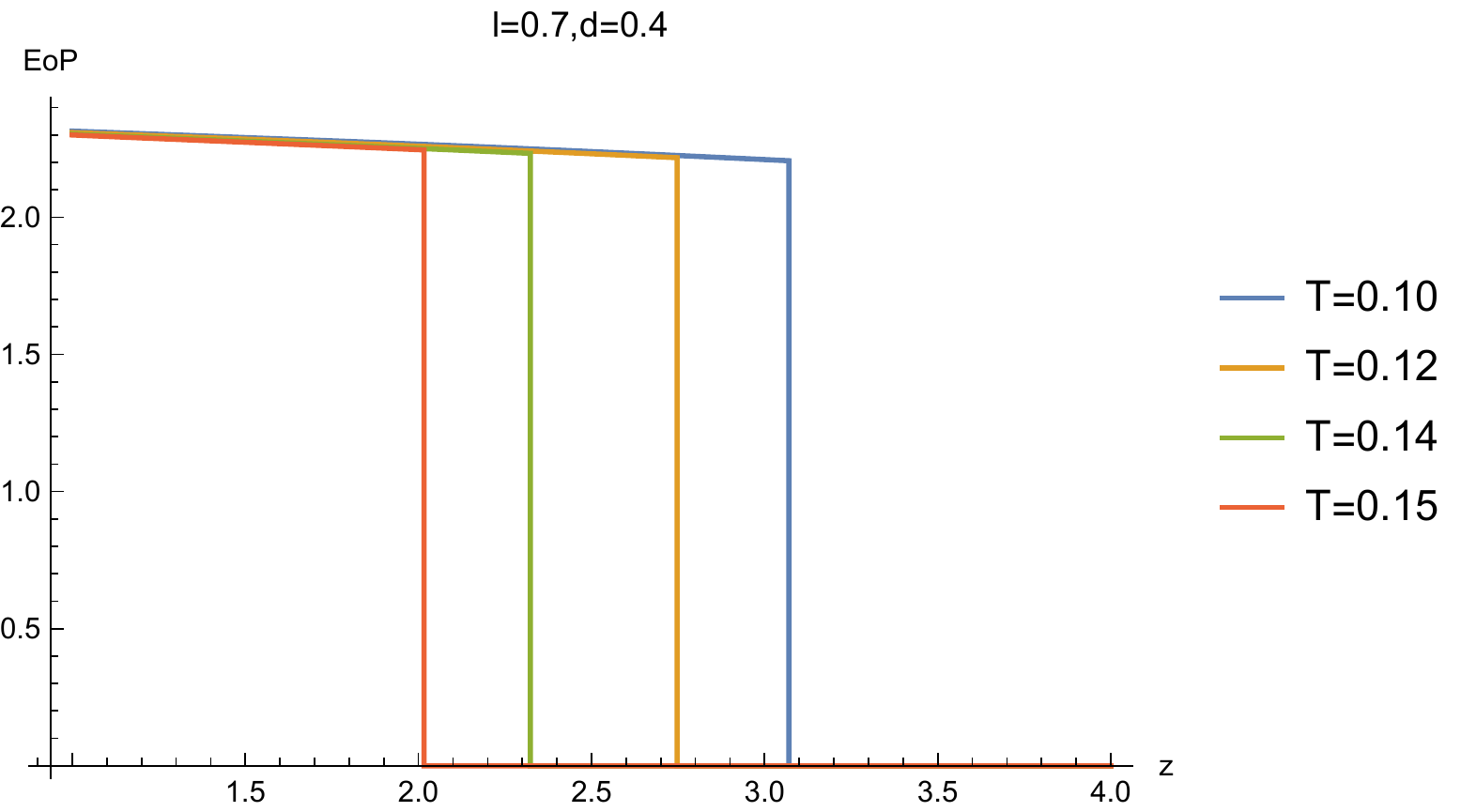}\ \hspace{0.2cm}
  \includegraphics[width=8cm]{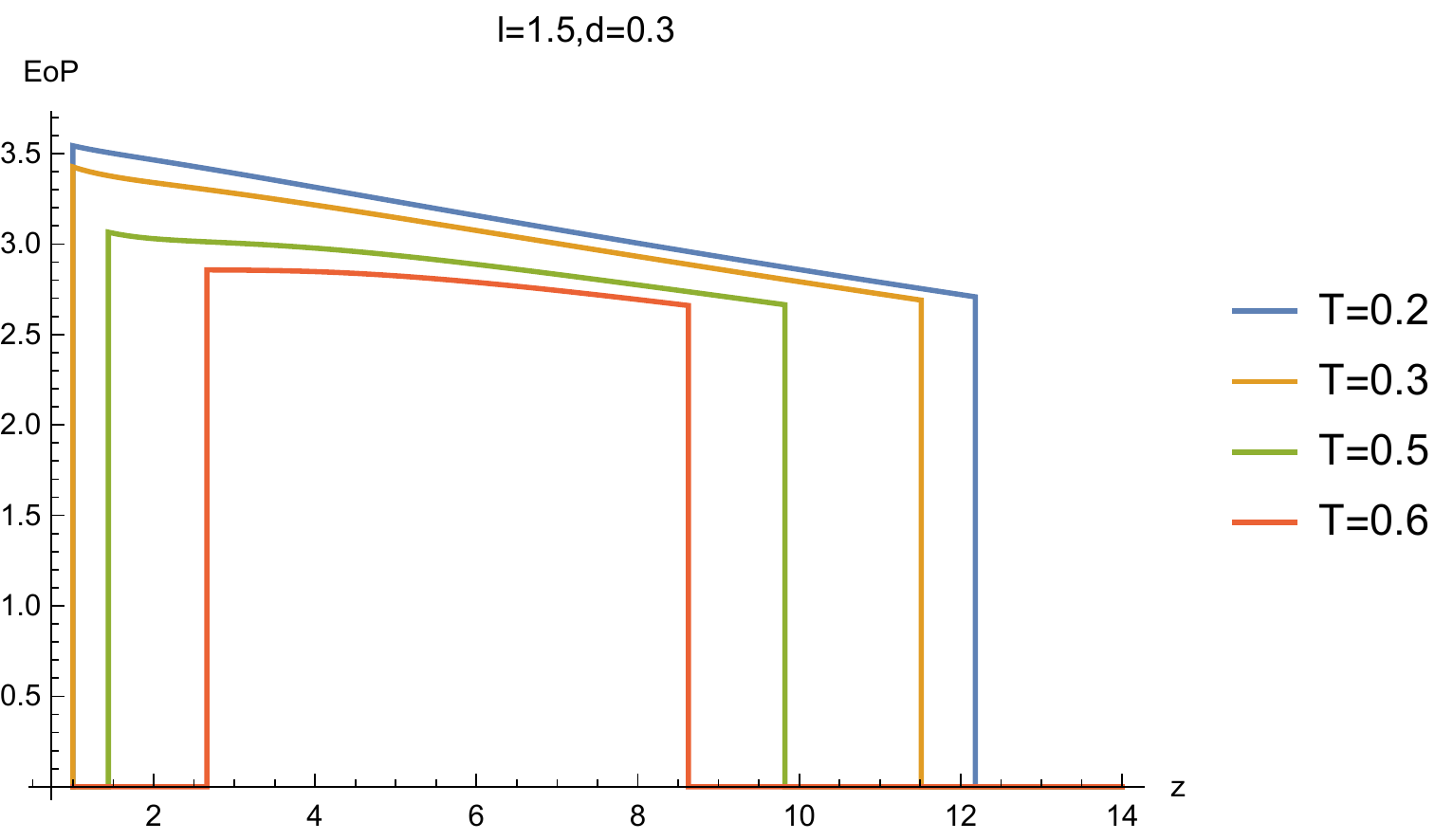}\ \\
  \caption{EoP as a function of $z$ for selected configuration parameters ($l$ and $d$) and temperatures.}
  \label{EoPvsz}
\end{figure}

%%%p
Now, we comment on how $z$ affects the EoP. There is no doubt that the critical lines of the EoP disentangling phase transition should be completely in agreement with those of MI. This is because when MI vanishes, the entanglement wedge is disconnected and so EoP is also zero. Furthermore, to compare the EoP with MI, we also plot EoP as the function $z$ for the same configuration parameters and temperatures as those of FIG.\ref{MIvsz}. We find that for $l=0.7$ and $d=0.4$ (see the left-hand plot in FIG. \ref{EoPvsz}), EoP slowly decreases as $z$ increases and no non-monotonic behavior emerges even for higher temperature. For the configuration parameters of $l=1.5$ and $d=0.3$ (see the right-hand plot in FIG. \ref{EoPvsz}), we also do not find obvious non-monotonic behavior. However, corresponding to the dome-shaped diagram in MI, a trapezoid-shaped diagram emerges in the diagram of EoP vs $z$ when the temperature becomes higher.

%%%%%

  Before closing this section, we argue that EoP can play a better role in characterizing the mixed state entanglement than HEE and MI. It is known that HEE can-not capture the mixed state entanglement well due to the fact that HEE takes into account both the quantum entanglement as well as the thermal effects for thermal states. MI, following the definition of HEE, can give a better diagnosis of the mixed state entanglement because MI can partly cancel the thermal effect from the HEE. However, this does not mean that MI can give a good diagnosis of the mixed state entanglement. As can be seen from FIG. \ref{HEE vs z}, the variation of HEE with $z$ always shows non-monotonic behavior. As a response, MI can also show a corresponding non-monotonic behavior (FIG. \ref{MIvsz}). Also, such an inverse non-monotonic behavior becomes more prominent with increasing temperature. The underlying reason comes from the dependency of MI on HEE, $I(A:C)=S_A+S_C-S_{A\cup C}$. When the temperature is large, the horizon radius becomes larger, and the minimum surface tends to approach the horizon and receive more contribution from the thermal entropy. As a result, $S_{A\cup C}$ can play a more dominant role in $I(A:C)$. Therefore, in this case, $I(A:C)$ can show an inverse non-monotonic behavior from that of HEE.

  From the analysis in the previous paragraph, we see that MI can be totally dominated by the thermal entropy. However, EoP does not suffer from this problem. A direct reason is that the minimal cross-section prescription of the holographic EoP includes the contribution from the entire bulk region, which will never be dictated by the near-horizon region. From FIG. \ref{EoPvsz} we see that the EoP always decreases with $z$ before hitting the critical point of disentangling phase transition, and no non-monotonic behavior following from the HEE can be observed. This evidence suggests the reliability of the independence of the EoP from the HEE and the thermal entropy. Combined with the observation in the previous paragraph, we can conclude that EoP can give a better diagnosis of the mixed entanglement measure than HEE and MI.

%%%%
\section{Discussion and conclusions}\label{section-e}

In this paper, some related informational quantities of holographic Lifshitz field theory have been numerically computed and their properties discussed in detail. These informational quantities reflect some universal entanglement properties of holographic relativistic dual field theory. These characteristics are summed up as:
%%%%
\begin{itemize}
  \item HEE decreases monotonically as the temperature falls, and finally approaches a finite value in the limit of zero temperature.
  \item MI decreases as the separation scale increases or the size of the subsystems decreases.
        Especially, a disentangling phase transition emerges as the separation scale increases or the subsystem size decreases.
        When we heat up the system, a disentangling phase transition in MI also occurs.
        These properties are universal and are independent of the configuration.
  \item The disentangling phase transition also emerges in EoP as the separation scale increases, the subsystem size decreases or the temperature rises. However, different from the case of MI, the change of EoP is abrupt, as it suddenly decreases to zero from a finite value.
\end{itemize}
%%%%%

The peculiar properties of the informational quantities of a holographic Lifshitz system are also explored.
An important property is the non-monotonicity of the HEE with $z$. The non-monotonicity emerges only for some specific configuration parameters and temperatures.

The non-monotonicity of the HEE with $z$ also leads to some non-monotonic behaviors in MI and EoP. Firstly, the disentangling phase transition point of MI and EoP as a function of $z$ is non-monotonic for some specific configurations and temperatures. Secondly, such non-monotonic behavior also emerges in MI. However, we note that for EoP, we cannot observe obvious non-monotonicity. In addition, we observe the novel phenomenon that a dome-shaped diagram emerges in MI vs $z$ for some configurations and temperatures. Correspondingly, a trapezoid-shaped diagram is also observed in EoP vs $z$. This means that for some specific configuration parameters and temperatures, the system measured in terms of MI and EoP is entangled only in some intermediate range of $z$.

Some open questions deserve further study.
\begin{itemize}
  \item An analytical study would surely provide more insights into our numerical results. Therefore, it would be valuable to study the related informational quantities in different regions analytically, especially at extreme temperature, or extreme system scale and separated scale, following Refs. \cite{Kundu:2016dyk,BabaeiVelni:2019pkw,Fischler:2012ca,Fischler:2012uv,Lala:2020lcp}.
  \item It is desirable to explore the non-equilibrium dynamics of these informational quantities in holographic Lifshitz dual field theory such that we can find more novel properties different from holographic relativistic dual field theory. Especially, it would be valuable to examine more inequalities of these related informational quantities in holographic non-equilibrium Lifshitz dual field theory. Some related topics have been explored, see Refs. \cite{BabaeiVelni:2020wfl,Zhou:2019xzc,Zhou:2019jlh} and references therein.
  \item Many interesting topics only focus on the HEE. For example, in Ref. \cite{Narayan:2012ks}, the authors study HEE in AdS plane waves, which pertain to certain hyperscaling-violating Lifshitz spacetimes and are dual to anisotropic excited systems. The HEE and Fisher information metric for a closed bosonic string in a homogeneous plane wave background are studied in Ref. \cite{Dimov:2017ryz}. Recently, the authors have further studied the HEE and Fisher information metric in Schroedinger spacetime.
      In Ref. \cite{Ghosh:2017ygi}, the authors study the deformation of the bulk minimal surface for both changes in the embeddings and the bulk metric. We can study mixed state entanglement, for example, MI and EoP, in these backgrounds.
  \item So far, there have been many studies of the informational quantities in quantum theory or experiments \cite{Casini:2009sr,Calabrese:2009qy,VanRaamsdonk:2016exw,Plenio:2007zz,Bennett:1996gf,Horodecki:2009zz}, including the informational quantities of several disjoint intervals, quantum quenches, entanglement purification protocols, etc. In particular, the von Neumann entropy, logarithmic negativity and odd entropy of Lifshitz scalar theories have been studied in Ref. \cite{MohammadiMozaffar:2017nri,MohammadiMozaffar:2017chk,MohammadiMozaffar:2018vmk,Mollabashi:2020ifv}. We can also study the theoretical information properties from the holographic side and see how well they match from the two sides.
\end{itemize}

%%%%%
\begin{acknowledgments}

  This work is supported by the Natural Science Foundation
  of China under Grants Nos. 11775036, 11905083, 11847055, 11705161 and Fok Ying Tung Education Foundation
  under Grant No. 171006. Guoyang Fu is supported by the Postgraduate Research \& Practice Innovation Program of Jiangsu Province (KYCX20\_2973). Jian-Pin Wu is also supported by Top Talent Support Program from Yangzhou University.

\end{acknowledgments}


\begin{thebibliography}{99}


  %%%%% AdS/CFT correspondence: Maldacena:1997re,Gubser:1998bc,Witten:1998qj,Aharony:1999ti

  %\cite{Maldacena:1997re}
  \bibitem{Maldacena:1997re}
  J.~M.~Maldacena,
  ``The Large N limit of superconformal field theories and supergravity,''
  Int.\ J.\ Theor.\ Phys.\  {\bf 38}, 1113 (1999)
  [Adv.\ Theor.\ Math.\ Phys.\  {\bf 2}, 231 (1998)]
  %doi:10.1023/A:1026654312961
  [hep-th/9711200].
  %%CITATION = doi:10.1023/A:1026654312961;%%
  %12024 citations counted in INSPIRE as of 14 Aug 2016

  %\cite{Gubser:1998bc}
  \bibitem{Gubser:1998bc}
  S.~S.~Gubser, I.~R.~Klebanov and A.~M.~Polyakov,
  ``Gauge theory correlators from noncritical string theory,''
  Phys.\ Lett.\ B {\bf 428}, 105 (1998)
  %doi:10.1016/S0370-2693(98)00377-3
  [hep-th/9802109].
  %%CITATION = doi:10.1016/S0370-2693(98)00377-3;%%
  %6851 citations counted in INSPIRE as of 14 Aug 2016

  %\cite{Witten:1998qj}
  \bibitem{Witten:1998qj}
  E.~Witten,
  ``Anti-de Sitter space and holography,''
  Adv.\ Theor.\ Math.\ Phys.\  {\bf 2}, 253 (1998)
  [hep-th/9802150].
  %%CITATION = HEP-TH/9802150;%%
  %7911 citations counted in INSPIRE as of 14 Aug 2016

  %\cite{Aharony:1999ti}
  \bibitem{Aharony:1999ti}
  O.~Aharony, S.~S.~Gubser, J.~M.~Maldacena, H.~Ooguri and Y.~Oz,
  ``Large N field theories, string theory and gravity,''
  Phys.\ Rept.\  {\bf 323}, 183 (2000)
  %doi:10.1016/S0370-1573(99)00083-6
  [hep-th/9905111].
  %%CITATION = doi:10.1016/S0370-1573(99)00083-6;%%
  %3816 citations counted in INSPIRE as of 14 Aug 2016

  %%%R-T formula: Ryu:2006bv,Takayanagi:2012kg,Lewkowycz:2013nqa

  %\cite{Ryu:2006bv}
  \bibitem{Ryu:2006bv}
  S.~Ryu and T.~Takayanagi,
  ``Holographic derivation of entanglement entropy from AdS/CFT,''
  Phys.\ Rev.\ Lett.\  {\bf 96}, 181602 (2006)
  %doi:10.1103/PhysRevLett.96.181602
  [hep-th/0603001].

  %\cite{Takayanagi:2012kg}
  \bibitem{Takayanagi:2012kg}
  T.~Takayanagi,
  ``Entanglement Entropy from a Holographic Viewpoint,''
  Class.\ Quant.\ Grav.\  {\bf 29}, 153001 (2012)
  %doi:10.1088/0264-9381/29/15/153001
  [arXiv:1204.2450 [gr-qc]].
  %%CITATION = doi:10.1088/0264-9381/29/15/153001;%%
  %168 citations counted in INSPIRE as of 07 Dec 2019

  %\cite{Lewkowycz:2013nqa}
  \bibitem{Lewkowycz:2013nqa}
  A.~Lewkowycz and J.~Maldacena,
  ``Generalized gravitational entropy,''
  JHEP {\bf 1308}, 090 (2013)
  %doi:10.1007/JHEP08(2013)090
  [arXiv:1304.4926 [hep-th]].

  %%%H-R-T formula:Hubeny:2007xt,Dong:2016hjy

  %\cite{Hubeny:2007xt}
  \bibitem{Hubeny:2007xt}
  V.~E.~Hubeny, M.~Rangamani and T.~Takayanagi,
  ``A Covariant holographic entanglement entropy proposal,''
  JHEP {\bf 0707}, 062 (2007)
  %doi:10.1088/1126-6708/2007/07/062
  [arXiv:0705.0016 [hep-th]].
  %%CITATION = doi:10.1088/1126-6708/2007/07/062;%%
  %880 citations counted in INSPIRE as of 07 Dec 2019

  %\cite{Dong:2016hjy}
  \bibitem{Dong:2016hjy}
  X.~Dong, A.~Lewkowycz and M.~Rangamani,
  ``Deriving covariant holographic entanglement,''
  JHEP {\bf 1611}, 028 (2016)
  %doi:10.1007/JHEP11(2016)028
  [arXiv:1607.07506 [hep-th]].

  %%%% HEE and QPT: Ling:2015dma,Ling:2016wyr,Ling:2016dck,Pakman:2008ui,Kuang:2014kha,Klebanov:2007ws,Zhang:2016rcm,Zeng:2016fsb
  %%% Thermodynamics: Albash:2012pd,Kuang:2014kha,Zhang:2016rcm,Zeng:2016fsb
  \bibitem{Albash:2012pd}
  T.~Albash and C.~V.~Johnson,
  ``Holographic Studies of Entanglement Entropy in Superconductors,''
  JHEP {\bf 1205}, 079 (2012)
  %doi:10.1007/JHEP05(2012)079
  [arXiv:1202.2605 [hep-th]].
  %%CITATION = doi:10.1007/JHEP05(2012)079;%%

  %\cite{Zhang:2016rcm}
  \bibitem{Zhang:2016rcm}
  S.~J.~Zhang,
  ``Holographic entanglement entropy close to crossover/phase transition in strongly coupled systems,''
  Nucl.\ Phys.\ B {\bf 916}, 304 (2017)
  %doi:10.1016/j.nuclphysb.2017.01.010
  [arXiv:1608.03072 [hep-th]].
  %%CITATION = doi:10.1016/j.nuclphysb.2017.01.010;%%
  %4 citations counted in INSPIRE as of 17 Mar 2019

  %\cite{Zeng:2016fsb}
  \bibitem{Zeng:2016fsb}
  X.~X.~Zeng and L.~F.~Li,
  ``Holographic Phase Transition Probed by Nonlocal Observables,''
  Adv.\ High Energy Phys.\  {\bf 2016}, 6153435 (2016)
  %doi:10.1155/2016/6153435
  [arXiv:1609.06535 [hep-th]].
  %%CITATION = doi:10.1155/2016/6153435;%%
  %23 citations counted in INSPIRE as of 17 Mar 2019

  %\cite{Kuang:2014kha}
  \bibitem{Kuang:2014kha}
  X.~M.~Kuang, E.~Papantonopoulos and B.~Wang,
  ``Entanglement Entropy as a Probe of the Proximity Effect in Holographic Superconductors,''
  JHEP {\bf 1405}, 130 (2014)
  [arXiv:1401.5720 [hep-th]].
  %%CITATION = ARXIV:1401.5720;%%
  %7 citations counted in INSPIRE as of 10 Feb 2015

  %%% QCP:   Ling:2015dma,Ling:2016wyr,Ling:2016dck,Pakman:2008ui,Klebanov:2007ws
  \bibitem{Ling:2015dma}
  Y.~Ling, {P.~Liu}, C.~Niu, J.~P.~Wu and Z.~Y.~Xian,
  ``{Holographic Entanglement Entropy Close to Quantum Phase Transitions,}''
  JHEP {\bf 1604}, 114 (2016)

  \bibitem{Ling:2016wyr}
  Y.~Ling, {P.~Liu} and J.~P.~Wu,
  ``{Characterization of Quantum Phase Transition using Holographic Entanglement Entropy,}''
  Phys.\ Rev.\ D {\bf 93}, no. 12, 126004 (2016)

  %\cite{Ling:2016dck}
  \bibitem{Ling:2016dck}
  Y.~Ling, P.~Liu, J.~P.~Wu and Z.~Zhou,
  ``Holographic Metal-Insulator Transition in Higher Derivative Gravity,''
  Phys.\ Lett.\ B {\bf 766}, 41 (2017)
  %doi:10.1016/j.physletb.2016.12.051
  [arXiv:1606.07866 [hep-th]].
  %%CITATION = doi:10.1016/j.physletb.2016.12.051;%%
  %18 citations counted in INSPIRE as of 17 Mar 2019

  %\cite{Pakman:2008ui}
  \bibitem{Pakman:2008ui}
  A.~Pakman and A.~Parnachev,
  ``Topological Entanglement Entropy and Holography,''
  JHEP {\bf 0807}, 097 (2008)
  % %%doi:10.1088/1126-6708/2008/07/097
  [arXiv:0805.1891 [hep-th]].
  %%CITATION = % %%doi:10.1088/1126-6708/2008/07/097;%%
  %54 citations counted in INSPIRE as of 24 Mar 2017

  %\cite{Klebanov:2007ws}
  \bibitem{Klebanov:2007ws}
  I.~R.~Klebanov, D.~Kutasov and A.~Murugan,
  ``Entanglement as a probe of confinement,''
  Nucl.\ Phys.\ B {\bf 796}, 274 (2008)
  % %%doi:10.1016/j.nuclphysb.2007.12.017
  [arXiv:0709.2140 [hep-th]].
  %%CITATION = % %%doi:10.1016/j.nuclphysb.2007.12.017;%%
  %213 citations counted in INSPIRE as of 24 Mar 2017

  %\cite{Arefeva:2020uec}
  \bibitem{Arefeva:2020uec}
  I.~Y.~Aref'eva, A.~Patrushev and P.~Slepov,
  ``Holographic entanglement entropy in anisotropic background with confinement-deconfinement phase transition,''
  JHEP \textbf{07} (2020), 043
  %doi:10.1007/JHEP07(2020)043
  [arXiv:2003.05847 [hep-th]].
  %1 citations counted in INSPIRE as of 01 Aug 2020

  %\cite{Liu:2020blk}
\bibitem{Liu:2020blk}
P.~Liu and J.~P.~Wu,
``Mixed State Entanglement and Thermal Phase Transitions,''
[arXiv:2009.01529 [hep-th]].
%0 citations counted in INSPIRE as of 16 Sep 2020

%%%%% Information theory:

\bibitem{Nielsen:QCQI}
M.~A.~Nielsen and I.~L.~Chuang, ``Quantum Computation and Quantum Information,''
Cambridge University Press.


  %\cite{Srednicki:1993im}
  \bibitem{Srednicki:1993im}
  M.~Srednicki,
  ``Entropy and area,''
  Phys. Rev. Lett. \textbf{71} (1993), 666-669
  %doi:10.1103/PhysRevLett.71.666
  [arXiv:hep-th/9303048 [hep-th]].
  %1138 citations counted in INSPIRE as of 04 Jul 2020

  %%%%: UV: Bombelli:1986,Srednicki:1993im
  \bibitem{Bombelli:1986}
  L. Bombelli, R. K. Koul, J. Lee and R. D. Sorkin, ``A Quantum Source of Entropy for Black
  Holes,'' Phys. Rev. D \textbf{34}, 373 (1986).

  %\cite{Casini:2004bw}
  \bibitem{Casini:2004bw}
  H.~Casini and M.~Huerta,
  ``A Finite entanglement entropy and the c-theorem,''
  Phys. Lett. B \textbf{600} (2004), 142-150
  %doi:10.1016/j.physletb.2004.08.072
  [arXiv:hep-th/0405111 [hep-th]].
  %246 citations counted in INSPIRE as of 04 Jul 2020

  %\cite{Headrick:2010zt}
  \bibitem{Headrick:2010zt}
  M.~Headrick,
  ``Entanglement Renyi entropies in holographic theories,''
  Phys.\ Rev.\ D {\bf 82}, 126010 (2010)
  %doi:10.1103/PhysRevD.82.126010
  [arXiv:1006.0047 [hep-th]].

  %\cite{Wolf:2007tdq}
  \bibitem{Wolf:2007tdq}
  M.~M.~Wolf, F.~Verstraete, M.~B.~Hastings and J.~I.~Cirac,
  ``Area Laws in Quantum Systems: Mutual Information and Correlations,''
  Phys.\ Rev.\ Lett.\  {\bf 100}, no. 7, 070502 (2008)
  %doi:10.1103/PhysRevLett.100.070502
  [arXiv:0704.3906 [quant-ph]].

  \bibitem{Fischler:2012uv}
  W.~Fischler, A.~Kundu and S.~Kundu,
  ``Holographic Mutual Information at Finite Temperature,''
  Phys.\ Rev.\ D {\bf 87}, no. 12, 126012 (2013)
  %doi:10.1103/PhysRevD.87.126012
  [arXiv:1212.4764 [hep-th]].

%%%% EOP in information theory: Terhal:0202044,Bagchi:1502

\bibitem{Terhal:0202044}
B. M. Terhal, M. Horodecki, D. W. Leung, D. P. DiVincenzo, ``The entanglement of purification,''
J. Math. Phys. 43, 4286--4298 (2002),
[arXiv:quant-ph/0202044].

\bibitem{Bagchi:1502}
S.~Bagchi, A. K. Pati, ``Monogamy, polygamy, and other properties of entanglement of purification,''
Phys. Rev. A 91, 042323 (2015),
[arXiv:1502.01272 [quant-ph]].

  %%%%% EOP: Takayanagi:2017knl,Nguyen:2017yqw

  %\cite{Takayanagi:2017knl}
  \bibitem{Takayanagi:2017knl}
  T.~Takayanagi and K.~Umemoto,
  ``Entanglement of purification through holographic duality,''
  Nature Phys.\  {\bf 14}, no. 6, 573 (2018)
  %doi:10.1038/s41567-018-0075-2
  [arXiv:1708.09393 [hep-th]].

  %\cite{Nguyen:2017yqw}
  \bibitem{Nguyen:2017yqw}
  P.~Nguyen, T.~Devakul, M.~G.~Halbasch, M.~P.~Zaletel and B.~Swingle,
  ``Entanglement of purification: from spin chains to holography,''
  JHEP {\bf 1801}, 098 (2018)
  %doi:10.1007/JHEP01(2018)098
  [arXiv:1709.07424 [hep-th]].

   %%%%: Bao:2018gck,Umemoto:2018jpc
  %40p
  %\cite{Bao:2018gck}
  \bibitem{Bao:2018gck}
  N.~Bao and I.~F.~Halpern,
  ``Conditional and Multipartite Entanglements of Purification and Holography,''
  Phys. Rev. D \textbf{99} (2019) no.4, 046010
  %doi:10.1103/PhysRevD.99.046010
  [arXiv:1805.00476 [hep-th]].
  %41 citations counted in INSPIRE as of 05 Jul 2020
  %41

  %\cite{Umemoto:2018jpc}
  \bibitem{Umemoto:2018jpc}
  K.~Umemoto and Y.~Zhou,
  ``Entanglement of Purification for Multipartite States and its Holographic Dual,''
  JHEP \textbf{10} (2018), 152
  %doi:10.1007/JHEP10(2018)152
  [arXiv:1805.02625 [hep-th]].
  %52 citations counted in INSPIRE as of 05 Jul 2020

  %\cite{Yang:2018gfq}
  \bibitem{Yang:2018gfq}
  R.~Q.~Yang, C.~Y.~Zhang and W.~M.~Li,
  ``Holographic entanglement of purification for thermofield double states and thermal quench,''
  JHEP \textbf{01} (2019), 114
  %doi:10.1007/JHEP01(2019)114
  [arXiv:1810.00420 [hep-th]].

  %%%%%% EoP for P. Liu: Liu:2019qje

  %\cite{Liu:2019qje}
  \bibitem{Liu:2019qje}
  P.~Liu, Y.~Ling, C.~Niu and J.~P.~Wu,
  ``Entanglement of Purification in Holographic Systems,''
  JHEP {\bf 1909}, 071 (2019)
  %doi:10.1007/JHEP09(2019)071
  [arXiv:1902.02243 [hep-th]].

  %\cite{Huang:2019zph}
  \bibitem{Huang:2019zph}
  Y.~f.~Huang, Z.~j.~Shi, C.~Niu, C.~y.~Zhang and P.~Liu,
  ``Mixed State Entanglement for Holographic Axion Model,''
  arXiv:1911.10977 [hep-th].
  %%CITATION = ARXIV:1911.10977;%%

  %\cite{Fu:2020oep}
  \bibitem{Fu:2020oep}
  G.~Fu, P.~Liu, H.~Gong, X.~M.~Kuang and J.~P.~Wu,
  ``Informational properties for Einstein-Maxwell-Dilaton Gravity,''
  [arXiv:2007.06001 [hep-th]].
  %0 citations counted in INSPIRE as of 19 Jul 2020

  %\cite{Ghodrati:2019hnn}
  \bibitem{Ghodrati:2019hnn}
  M.~Ghodrati, X.~M.~Kuang, B.~Wang, C.~Y.~Zhang and Y.~T.~Zhou,
  ``The connection between holographic entanglement and complexity of purification,''
  JHEP {\bf 1909} (2019) 009
  %doi:10.1007/JHEP09(2019)009
  [arXiv:1902.02475 [hep-th]].
  %%CITATION = doi:10.1007/JHEP09(2019)009;%%
  %16 citations counted in INSPIRE as of 24 Feb 2020

  %%% Gravity Lifshitz

\bibitem{Hohenberg:1977}
P. C. Hohenberg and B. I. Halperin, ``Theory of Dynamic Critical Phenomena,'' Rev. Mod. Phys. {\bf 49} (1977) 435.

%%%%%: Natsuume:2018yrg,Li:2019oyz

%\cite{Natsuume:2018yrg}
\bibitem{Natsuume:2018yrg}
M.~Natsuume and T.~Okamura,
``Holographic Lifshitz superconductors: Analytic solution,''
Phys. Rev. D \textbf{97} (2018) no.6, 066016
%doi:10.1103/PhysRevD.97.066016
[arXiv:1801.03154 [hep-th]].
%4 citations counted in INSPIRE as of 23 Jan 2021

%\cite{Li:2019oyz}
\bibitem{Li:2019oyz}
Z.~H.~Li, C.~Y.~Xia, H.~B.~Zeng and H.~Q.~Zhang,
``Formation and critical dynamics of topological defects in Lifshitz holography,''
JHEP \textbf{04} (2020), 147
%doi:10.1007/JHEP04(2020)147
[arXiv:1912.10450 [hep-th]].
%2 citations counted in INSPIRE as of 23 Jan 2021

  %\cite{Hertz:1976zz}
  \bibitem{Hertz:1976zz}
  J.~A.~Hertz,
  ``Quantum critical phenomena,''
  Phys.\ Rev.\ B {\bf 14}, 1165 (1976).
  %doi:10.1103/PhysRevB.14.1165

  %%%%: Lifshitz black hole: Kachru:2008yh,Danielsson:2009gi,Mann:2009yx,Bertoldi:2009vn,Taylor:2008tg,Pang:2009ad,Pang:2009pd,Balasubramanian:2009rx,AyonBeato:2009nh,
  % Cai:2009ac,Myung:2009up,Dehghani:2011tx,Keranen:2012mx,Tarrio:2011de,Kuang:2017rpx

  %\cite{Kachru:2008yh}
  \bibitem{Kachru:2008yh}
  S.~Kachru, X.~Liu and M.~Mulligan,
  ``Gravity duals of Lifshitz-like fixed points,''
  Phys.\ Rev.\ D {\bf 78}, 106005 (2008)
  %doi:10.1103/PhysRevD.78.106005
  [arXiv:0808.1725 [hep-th]].

  \bibitem{Danielsson:2009gi}
  U.~H.~Danielsson and L.~Thorlacius,
  ``Black holes in asymptotically Lifshitz spacetime,''
  JHEP {\bf 0903}, 070 (2009)
  %doi:10.1088/1126-6708/2009/03/070
  [arXiv:0812.5088 [hep-th]].

  \bibitem{Mann:2009yx}
  R.~B.~Mann,
  ``Lifshitz Topological Black Holes,''
  JHEP {\bf 0906}, 075 (2009)
  %doi:10.1088/1126-6708/2009/06/075
  [arXiv:0905.1136 [hep-th]].

  %\cite{Bertoldi:2009vn}
  \bibitem{Bertoldi:2009vn}
  G.~Bertoldi, B.~A.~Burrington and A.~Peet,
  ``Black Holes in asymptotically Lifshitz spacetimes with arbitrary critical exponent,''
  Phys.\ Rev.\ D {\bf 80}, 126003 (2009)
  %doi:10.1103/PhysRevD.80.126003
  [arXiv:0905.3183 [hep-th]].

  \bibitem{Taylor:2008tg}
  M.~Taylor,
  ``Non-relativistic holography,''
  arXiv:0812.0530 [hep-th].

  %\cite{Pang:2009ad}
  \bibitem{Pang:2009ad}
  D.~W.~Pang,
  ``A Note on Black Holes in Asymptotically Lifshitz Spacetime,''
  Commun.\ Theor.\ Phys.\  {\bf 62}, 265 (2014)
  % doi:10.1088/0253-6102/62/2/14
  [arXiv:0905.2678 [hep-th]].

  %\cite{Pang:2009pd}
  \bibitem{Pang:2009pd}
  D.~W.~Pang,
  ``On Charged Lifshitz Black Holes,''
  JHEP {\bf 1001}, 116 (2010)
  %doi:10.1007/JHEP01(2010)116
  [arXiv:0911.2777 [hep-th]].

  %\cite{Balasubramanian:2009rx}
  \bibitem{Balasubramanian:2009rx}
  K.~Balasubramanian and J.~McGreevy,
  ``An Analytic Lifshitz black hole,''
  Phys.\ Rev.\ D {\bf 80}, 104039 (2009)
  %doi:10.1103/PhysRevD.80.104039
  [arXiv:0909.0263 [hep-th]].

  %\cite{AyonBeato:2009nh}
  \bibitem{AyonBeato:2009nh}
  E.~Ayon-Beato, A.~Garbarz, G.~Giribet and M.~Hassaine,
  ``Lifshitz Black Hole in Three Dimensions,''
  Phys.\ Rev.\ D {\bf 80}, 104029 (2009)
  %doi:10.1103/PhysRevD.80.104029
  [arXiv:0909.1347 [hep-th]].

  %\cite{Cai:2009ac}
  \bibitem{Cai:2009ac}
  R.~G.~Cai, Y.~Liu and Y.~W.~Sun,
  ``A Lifshitz Black Hole in Four Dimensional R**2 Gravity,''
  JHEP {\bf 0910}, 080 (2009)
  %doi:10.1088/1126-6708/2009/10/080
  [arXiv:0909.2807 [hep-th]].

  %\cite{Myung:2009up}
  \bibitem{Myung:2009up}
  Y.~S.~Myung, Y.~W.~Kim and Y.~J.~Park,
  ``Dilaton gravity approach to three dimensional Lifshitz black hole,''
  Eur.\ Phys.\ J.\ C {\bf 70}, 335 (2010)
  %doi:10.1140/epjc/s10052-010-1460-x
  [arXiv:0910.4428 [hep-th]].

  %\cite{Dehghani:2011tx}
  \bibitem{Dehghani:2011tx}
  M.~H.~Dehghani, R.~B.~Mann and R.~Pourhasan,
  ``Charged Lifshitz Black Holes,''
  Phys.\ Rev.\ D {\bf 84}, 046002 (2011)
  % doi:10.1103/PhysRevD.84.046002
  [arXiv:1102.0578 [hep-th]].

  %\cite{Keranen:2012mx}
  \bibitem{Keranen:2012mx}
  V.~Keranen and L.~Thorlacius,
  ``Thermal Correlators in Holographic Models with Lifshitz scaling,''
  Class.\ Quant.\ Grav.\  {\bf 29}, 194009 (2012)
  %doi:10.1088/0264-9381/29/19/194009
  [arXiv:1204.0360 [hep-th]].

  %\cite{Tarrio:2011de}
  \bibitem{Tarrio:2011de}
  J.~Tarrio and S.~Vandoren,
  ``Black holes and black branes in Lifshitz spacetimes,''
  JHEP {\bf 1109}, 017 (2011)
  %doi:10.1007/JHEP09(2011)017
  [arXiv:1105.6335 [hep-th]].

  %\cite{Kuang:2017rpx}
  \bibitem{Kuang:2017rpx}
  X.~M.~Kuang, E.~Papantonopoulos, J.~P.~Wu and Z.~Zhou,
  ``Lifshitz black branes and DC transport coefficients in massive Einstein-Maxwell-dilaton gravity,''
  Phys.\ Rev.\ D {\bf 97}, no. 6, 066006 (2018)
  %doi:10.1103/PhysRevD.97.066006
  [arXiv:1709.02976 [hep-th]].

  %\cite{MolinaVilaplana:2011xt}
  \bibitem{MolinaVilaplana:2011xt}
  J.~Molina-Vilaplana and P.~Sodano,
  ``Holographic View on Quantum Correlations and Mutual Information between Disjoint Blocks of a Quantum Critical System,''
  JHEP {\bf 1110}, 011 (2011)
  %doi:10.1007/JHEP10(2011)011
  [arXiv:1108.1277 [quant-ph]].

  %%% HEE with Lifshitz: Chakraborty:2014lfa,Karar:2020cvz

  %\cite{Chakraborty:2014lfa}
  \bibitem{Chakraborty:2014lfa}
  S.~Chakraborty, P.~Dey, S.~Karar and S.~Roy,
  ``Entanglement thermodynamics for an excited state of Lifshitz system,''
  JHEP \textbf{04} (2015), 133
  %doi:10.1007/JHEP04(2015)133
  [arXiv:1412.1276 [hep-th]].
  %14 citations counted in INSPIRE as of 23 Jul 2020

  %\cite{Karar:2020cvz}
  \bibitem{Karar:2020cvz}
  S.~Karar and S.~Gangopadhyay,
  ``Holographic information theoretic quantities for Lifshitz black hole,''
  Eur. Phys. J. C \textbf{80} (2020) no.6, 515
  %doi:10.1140/epjc/s10052-020-8091-7
  [arXiv:2002.08272 [hep-th]].
  %0 citations counted in INSPIRE as of 23 Jul 2020

  %\cite{Dong:2012se}
  \bibitem{Dong:2012se}
  X.~Dong, S.~Harrison, S.~Kachru, G.~Torroba and H.~Wang,
  ``Aspects of holography for theories with hyperscaling violation,''
  JHEP \textbf{06} (2012), 041
  %doi:10.1007/JHEP06(2012)041
  [arXiv:1201.1905 [hep-th]].
  %276 citations counted in INSPIRE as of 23 Jul 2020

  %\cite{Alishahiha:2015goa}
  \bibitem{Alishahiha:2015goa}
  M.~Alishahiha, A.~F.~Astaneh, P.~Fonda and F.~Omidi,
  ``Entanglement Entropy for Singular Surfaces in Hyperscaling violating Theories,''
  JHEP \textbf{09} (2015), 172
  %doi:10.1007/JHEP09(2015)172
  [arXiv:1507.05897 [hep-th]].
  %22 citations counted in INSPIRE as of 23 Jul 2020

  %%%p: Mishra:2018tzj,
  %\cite{Mishra:2018tzj}
  \bibitem{Mishra:2018tzj}
  R.~Mishra and H.~Singh,
  ``Entanglement entropy at higher orders for the states of $a = 3$ Lifshitz theory,''
  Nucl. Phys. B \textbf{938} (2019), 307-320
  %doi:10.1016/j.nuclphysb.2018.11.012
  [arXiv:1804.01361 [hep-th]].
%5 citations counted in INSPIRE as of 16 Sep 2020

  %\cite{Wu:2013xta}
  \bibitem{Wu:2013xta}
  J.~P.~Wu,
  ``Holographic fermions on a charged Lifshitz background from Einstein-Dilaton-Maxwell model,''
  JHEP \textbf{03} (2013), 083
  %doi:10.1007/JHEP03(2013)083
  %11 citations counted in INSPIRE as of 25 Jul 2020

  %\cite{Wu:2014rqa}
  \bibitem{Wu:2014rqa}
  J.~P.~Wu,
  ``The charged Lifshitz black brane geometry and the bulk dipole coupling,''
  Phys. Lett. B \textbf{728} (2014), 450-456
  %doi:10.1016/j.physletb.2013.11.040
  %16 citations counted in INSPIRE as of 25 Jul 2020

%%%%%:  Solodukhin:2009sk,Nesterov:2010yi

%\cite{Solodukhin:2009sk}
\bibitem{Solodukhin:2009sk}
S.~N.~Solodukhin,
``Entanglement Entropy in Non-Relativistic Field Theories,''
JHEP \textbf{04} (2010), 101
%doi:10.1007/JHEP04(2010)101
[arXiv:0909.0277 [hep-th]].
%30 citations counted in INSPIRE as of 27 Jan 2021

  %\cite{Nesterov:2010yi}
\bibitem{Nesterov:2010yi}
D.~Nesterov and S.~N.~Solodukhin,
``Gravitational effective action and entanglement entropy in UV modified theories with and without Lorentz symmetry,''
Nucl. Phys. B \textbf{842} (2011), 141-171
%doi:10.1016/j.nuclphysb.2010.08.006
[arXiv:1007.1246 [hep-th]].
%43 citations counted in INSPIRE as of 27 Jan 2021

%\cite{Gentle:2015cfp}
\bibitem{Gentle:2015cfp}
S.~A.~Gentle and C.~Keeler,
``On the reconstruction of Lifshitz spacetimes,''
JHEP \textbf{03} (2016), 195
%doi:10.1007/JHEP03(2016)195
[arXiv:1512.04538 [hep-th]].
%19 citations counted in INSPIRE as of 27 Jan 2021

%\cite{Cheyne:2017bis}
\bibitem{Cheyne:2017bis}
J.~Cheyne and D.~Mattingly,
``Constructing entanglement wedges for Lifshitz spacetimes with Lifshitz gravity,''
Phys. Rev. D \textbf{97} (2018) no.6, 066024
%doi:10.1103/PhysRevD.97.066024
[arXiv:1707.05913 [gr-qc]].
%8 citations counted in INSPIRE as of 27 Jan 2021

%\cite{Janiszewski:2017tas}
\bibitem{Janiszewski:2017tas}
S.~Janiszewski,
``Non-relativistic entanglement entropy from Horava gravity,''
[arXiv:1707.08231 [hep-th]].
%1 citations counted in INSPIRE as of 27 Jan 2021

  %%%:   Kundu:2016dyk,BabaeiVelni:2019pkw,Fischler:2012ca,Fischler:2012uv
  %\cite{Kundu:2016dyk}
  \bibitem{Kundu:2016dyk}
  S.~Kundu and J.~F.~Pedraza,
  ``Aspects of Holographic Entanglement at Finite Temperature and Chemical Potential,''
  JHEP \textbf{08} (2016), 177
  %doi:10.1007/JHEP08(2016)177
  [arXiv:1602.07353 [hep-th]].
  %42 citations counted in INSPIRE as of 27 Jul 2020

  %\cite{BabaeiVelni:2019pkw}
  \bibitem{BabaeiVelni:2019pkw}
  K.~Babaei Velni, M.~R.~Mohammadi Mozaffar and M.~H.~Vahidinia,
  ``Some Aspects of Entanglement Wedge Cross-Section,''
  JHEP \textbf{05} (2019), 200
  %doi:10.1007/JHEP05(2019)200
  [arXiv:1903.08490 [hep-th]].
  %18 citations counted in INSPIRE as of 27 Jul 2020

  %\cite{Fischler:2012ca}
  \bibitem{Fischler:2012ca}
  W.~Fischler and S.~Kundu,
  ``Strongly Coupled Gauge Theories: High and Low Temperature Behavior of Non-local Observables,''
  JHEP \textbf{05} (2013), 098
  %doi:10.1007/JHEP05(2013)098
  [arXiv:1212.2643 [hep-th]].
  %76 citations counted in INSPIRE as of 27 Jul 2020

  %\cite{Lala:2020lcp}
  \bibitem{Lala:2020lcp}
  A.~Lala,
  ``Entanglement measures for non-conformal D-branes,''
  [arXiv:2008.06154 [hep-th]].
  %1 citations counted in INSPIRE as of 16 Sep 2020

  %%% :BabaeiVelni:2020wfl,Zhou:2019xzc,Zhou:2019jlh

  %\cite{BabaeiVelni:2020wfl}
  \bibitem{BabaeiVelni:2020wfl}
  K.~Babaei Velni, M.~R.~Mohammadi Mozaffar and M.~H.~Vahidinia,
  ``Evolution of Entanglement Wedge Cross Section Following a Global Quench,''
  [arXiv:2005.05673 [hep-th]].
  %1 citations counted in INSPIRE as of 27 Jul 2020

  %\cite{Zhou:2019xzc}
  \bibitem{Zhou:2019xzc}
  Y.~T.~Zhou, X.~M.~Kuang, Y.~Z.~Li and J.~P.~Wu,
  ``Holographic subregion complexity under a thermal quench in an Einstein-Maxwell-axion theory with momentum relaxation,''
  Phys. Rev. D \textbf{101} (2020) no.10, 106024
  %doi:10.1103/PhysRevD.101.106024
  [arXiv:1912.03479 [hep-th]].
  %0 citations counted in INSPIRE as of 27 Jul 2020

  %\cite{Zhou:2019jlh}
  \bibitem{Zhou:2019jlh}
  Y.~T.~Zhou, M.~Ghodrati, X.~M.~Kuang and J.~P.~Wu,
  ``Evolutions of entanglement and complexity after a thermal quench in massive gravity theory,''
  Phys. Rev. D \textbf{100} (2019) no.6, 066003
  %doi:10.1103/PhysRevD.100.066003
  [arXiv:1907.08453 [hep-th]].
  %11 citations counted in INSPIRE as of 27 Jul 2020

  %\cite{Narayan:2012ks}
\bibitem{Narayan:2012ks}
K.~Narayan, T.~Takayanagi and S.~P.~Trivedi,
``AdS plane waves and entanglement entropy,''
JHEP \textbf{04} (2013), 051
%doi:10.1007/JHEP04(2013)051
[arXiv:1212.4328 [hep-th]].
%36 citations counted in INSPIRE as of 16 Sep 2020

%\cite{Dimov:2017ryz}
\bibitem{Dimov:2017ryz}
H.~Dimov, S.~Mladenov, R.~C.~Rashkov and T.~Vetsov,
``Entanglement entropy and Fisher information metric for closed bosonic strings in homogeneous plane wave background,''
Phys. Rev. D \textbf{96} (2017) no.12, 126004
%doi:10.1103/PhysRevD.96.126004
[arXiv:1705.01873 [hep-th]].
%8 citations counted in INSPIRE as of 16 Sep 2020

%\cite{Ghosh:2017ygi}
\bibitem{Ghosh:2017ygi}
A.~Ghosh and R.~Mishra,
``Inhomogeneous Jacobi equation for minimal surfaces and perturbative change in holographic entanglement entropy,''
Phys. Rev. D \textbf{97} (2018) no.8, 086012
%doi:10.1103/PhysRevD.97.086012
[arXiv:1710.02088 [hep-th]].
%12 citations counted in INSPIRE as of 16 Sep 2020



%%%Hua-jie Gong add

%\cite{Casini:2009sr}
\bibitem{Casini:2009sr}
H.~Casini and M.~Huerta,
``Entanglement entropy in free quantum field theory,''
J. Phys. A \textbf{42} (2009), 504007
%doi:10.1088/1751-8113/42/50/504007
[arXiv:0905.2562 [hep-th]].
%446 citations counted in INSPIRE as of 28 Feb 2021

%\cite{Calabrese:2009qy}
\bibitem{Calabrese:2009qy}
P.~Calabrese and J.~Cardy,
``Entanglement entropy and conformal field theory,''
J. Phys. A \textbf{42} (2009), 504005
%doi:10.1088/1751-8113/42/50/504005
[arXiv:0905.4013 [cond-mat.stat-mech]].
%697 citations counted in INSPIRE as of 28 Feb 2021

%\cite{VanRaamsdonk:2016exw}
\bibitem{VanRaamsdonk:2016exw}
M.~Van Raamsdonk,
``Lectures on Gravity and Entanglement,''
%doi:10.1142/9789813149441\_0005
[arXiv:1609.00026 [hep-th]].
%67 citations counted in INSPIRE as of 28 Feb 2021

%\cite{Plenio:2007zz}
\bibitem{Plenio:2007zz}
M.~B.~Plenio and S.~Virmani,
``An Introduction to entanglement measures,''
Quant. Inf. Comput. \textbf{7} (2007), 1-51
[arXiv:quant-ph/0504163 [quant-ph]].
%124 citations counted in INSPIRE as of 28 Feb 2021

%\cite{Bennett:1996gf}
\bibitem{Bennett:1996gf}
C.~H.~Bennett, D.~P.~DiVincenzo, J.~A.~Smolin and W.~K.~Wootters,
``Mixed state entanglement and quantum error correction,''
Phys. Rev. A \textbf{54} (1996), 3824-3851
%doi:10.1103/PhysRevA.54.3824
[arXiv:quant-ph/9604024 [quant-ph]].
%390 citations counted in INSPIRE as of 28 Feb 2021

%\cite{Horodecki:2009zz}
\bibitem{Horodecki:2009zz}
R.~Horodecki, P.~Horodecki, M.~Horodecki and K.~Horodecki,
``Quantum entanglement,''
Rev. Mod. Phys. \textbf{81} (2009), 865-942
%doi:10.1103/RevModPhys.81.865
[arXiv:quant-ph/0702225 [quant-ph]].
%587 citations counted in INSPIRE as of 28 Feb 2021

%%%End


%%%%: MohammadiMozaffar:2017nri,MohammadiMozaffar:2017chk,MohammadiMozaffar:2018vmk,Mollabashi:2020ifv

%\cite{MohammadiMozaffar:2017nri}
\bibitem{MohammadiMozaffar:2017nri}
M.~R.~Mohammadi Mozaffar and A.~Mollabashi,
``Entanglement in Lifshitz-type Quantum Field Theories,''
JHEP \textbf{07} (2017), 120
%doi:10.1007/JHEP07(2017)120
[arXiv:1705.00483 [hep-th]].
%31 citations counted in INSPIRE as of 16 Sep 2020

%\cite{MohammadiMozaffar:2017chk}
\bibitem{MohammadiMozaffar:2017chk}
M.~R.~Mohammadi Mozaffar and A.~Mollabashi,
``Logarithmic Negativity in Lifshitz Harmonic Models,''
J. Stat. Mech. \textbf{1805} (2018) no.5, 053113
%doi:10.1088/1742-5468/aac135
[arXiv:1712.03731 [hep-th]].
%11 citations counted in INSPIRE as of 16 Sep 2020

%\cite{MohammadiMozaffar:2018vmk}
\bibitem{MohammadiMozaffar:2018vmk}
M.~R.~Mohammadi Mozaffar and A.~Mollabashi,
``Entanglement Evolution in Lifshitz-type Scalar Theories,''
JHEP \textbf{01} (2019), 137
%doi:10.1007/JHEP01(2019)137
[arXiv:1811.11470 [hep-th]].
%12 citations counted in INSPIRE as of 16 Sep 2020

%\cite{Mollabashi:2020ifv}
\bibitem{Mollabashi:2020ifv}
A.~Mollabashi and K.~Tamaoka,
``A Field Theory Study of Entanglement Wedge Cross Section: Odd Entropy,''
JHEP \textbf{08} (2020), 078
%doi:10.1007/JHEP08(2020)078
[arXiv:2004.04163 [hep-th]].
%3 citations counted in INSPIRE as of 16 Sep 2020


\end{thebibliography}
\end{document}